%% file: main_epjc.tex
\documentclass[twocolumn,epj,nopacs]{svjour}  
\smartqed  
\usepackage[switch]{lineno}

\usepackage{amsfonts}
\usepackage{amssymb}
\usepackage{amsmath}
\usepackage[titletoc]{appendix}
\usepackage{cite}
\usepackage{cancel}
\usepackage{soul}
\usepackage{graphicx} 
\usepackage{caption,subcaption}

\usepackage{dcolumn}
\usepackage{bm}
\usepackage{slashed}

\usepackage{IEEEtrantools}
\usepackage{csquotes}
\usepackage{makecell}
\usepackage{color}
\usepackage[version=4]{mhchem}

\usepackage{comment}
\usepackage{multirow}
\usepackage{soul}

\usepackage[colorlinks = true,
            linkcolor = blue,
            urlcolor  = blue,
            citecolor = blue,
            anchorcolor = blue]{hyperref}
\usepackage{tikz,xcolor}

\journalname{Prepared for EPJC}

\usepackage{enumitem}
\newlist{enumcompactenum}{enumerate}{3}
\setlist[enumcompactenum]{topsep=0pt,partopsep=0pt,itemsep=0pt,parsep=0pt}
\setlist[enumcompactenum,1]{label=\arabic*}
\setlist[enumcompactenum,2]{label=\alph*}
\setlist[enumcompactenum,3]{label=\roman*}

\usepackage{xcolor}

\begin{document}

\titlerunning{Bayesian network 3D event reconstruction in the Cygno TPC}

\input{author_list_EPJC-new}

\date{\today}

\abstract{
The CYGNO experiment is developing a high-resolution gaseous Time Projection Chamber with optical readout for directional dark matter searches. The detector uses a helium-tetrafluoromethane (He:CF$_4$ 60:40) gas mixture at atmospheric pressure and a triple Gas Electron Multiplier amplification stage, coupled with a scientific camera for high-resolution 2D imaging and fast photomultipliers for time-resolved scintillation light detection. This setup enables 3D event reconstruction: photomultiplier signals provide depth information, while the camera delivers high-precision transverse resolution. In this work, we present a Bayesian Network-based algorithm designed to reconstruct the events using only the photomultiplier signals, inferring a 3D description of the particle trajectories. The algorithm models the light collection process probabilistically and estimates spatial and intensity parameters on the Gas Electron Multiplier plane, where light emission occurs. It is implemented within the Bayesian Analysis Toolkit and uses Markov Chain Monte Carlo sampling for posterior inference. Validation using data from the CYGNO LIME prototype shows accurate reconstruction of localized and extended straight tracks. Results demonstrate that the Bayesian approach enables robust 3D description and, when combined with camera data, opens the way to future improvements in spatial and energy resolution. This methodology represents a significant step forward in directional dark matter detection, enhancing the identification of nuclear recoil tracks with high spatial resolution.}

\title{Bayesian network 3D event reconstruction in the Cygno optical TPC for dark matter direct detection}
\maketitle

\section{Introduction} \label{sec:intro}
Direct detection of dark matter (DM) remains one of the most significant challenges in modern physics. Although astrophysical and cosmological observations provide compelling evidence for its existence~\cite{Bertone:2004pz}, direct interactions between dark matter and ordinary matter remain unconfirmed~\cite{ParticleDataGroup:2024cfk, XENON:2016jmt, XENON:2019gfn, DarkSide-50:2023fcw, DarkSide:2018bpj, DarkSide:2018kuk, DarkSide:2018ppu, XENON:2019zpr, XENONCollaboration:2023dar, SuperCDMS:2024yiv, DEAPCollaboration:2021raj}. Among the most promising approaches is directional detection, which seeks to identify the characteristic anisotropy of dark matter-induced nuclear recoils, expected to align with the Solar System’s motion through the Galactic halo~\cite{Mayet_2016zxu}. High-resolution 3D tracking of particle interactions would greatly facilitate achieving this goal~\cite{Green:2006cb}, enabling us to resolve sub-millimeter structures in low-energy events.

The CYGNO experiment~\cite{Amaro:2022gub}, part of the international CYGNUS proto-collaboration~\cite{Vahsen:2020pzb}, is developing a high-resolution gaseous Time Projection Chamber (TPC) with optical readout, optimized for directional dark matter detection~\cite{Deaconu:2017vam}. The TPC uses a helium-tetrafluoromethane (He:CF$_4$) gas mixture in the 60:40 ratio at atmospheric pressure, which allows for efficient ionization and scintillation~\cite{bib:Fraga:2003uu, Antochi:2018otx, Miernik:2007cwz}. Charged particles interacting with the gas create ionization tracks. The resulting electrons drift under a uniform electric field toward a triple Gas Electron Multiplier (GEM)~\cite{Sauli1997} stage, where they are amplified and generate secondary light emission. This light is recorded by two complementary detection systems: an Active Pixel Sensor of type scientific Complementary MOS (APS-sCMOS), which captures a high-resolution 2D projection of the event on the GEM plane, and photomultipliers (PMTs), which collect time-resolved scintillation light, providing information on the particle’s path along the longitudinal (drift) coordinate. Although Timepix-based cameras could in principle provide nanosecond time resolution and hence a 3D reconstruction, they offer a much smaller number of pixels, higher per-pixel noise, and significantly higher costs. For large-area optical readouts, the APS-sCMOS combined with PMTs, as adopted in this work, remains more suitable.

While the camera provides detailed spatial information in the plane parallel to the GEM stack ($XY$), it lacks direct depth sensitivity, making it inherently a 2D imaging system. In contrast, PMTs provide time-resolved signals of the light emitted, allowing the reconstruction of the development of the track along the direction orthogonal to the GEM plane ($\Delta Z$). Moreover, since the intensity of the light collected by the PMTs depends on the emission point on the GEM plane, it is possible to infer the transverse ($XY$) position of the source as well. Therefore, by analyzing the PMT signals, a 3D event reconstruction can be achieved, independent of the camera image.

In this work we focus on the 3D reconstruction of short straight tracks, which are the expected signature of low-energy nuclear recoils induced by dark matter interactions. Our present goal is to demonstrate robust reconstruction of this topology, which provides the basis for future extraction of the track angle and, ultimately, the direction of the incoming particle. Such directionality would offer a powerful handle to discriminate dark matter from background. For this reason, the reconstruction algorithm is optimized for straight-track topologies. More sophisticated approaches could in principle be developed to recover more complex event shapes, such as those induced by longer electronic recoils, but this lies beyond the scope of the present work and represents a promising direction for future studies. 

Several collaborations pursue the development of gaseous TPCs 
for directional detection, such as 
NEWAGE~\cite{NEWAGE2015}, 
MIMAC~\cite{MIMAC2011}, 
DMTPC~\cite{DMTPC2011}, 
DRIFT~\cite{DRIFT2012}, 
all aiming to exploit track reconstruction to achieve sensitivity to the incoming WIMP direction. 
In addition to gaseous TPCs, R\&D efforts have also explored the possibility of exploiting 
columnar recombination in liquid argon to gain directional sensitivity, as pursued in the RED 
program associated with DarkSide~\cite{DarkSide20k:2025tes,Bondar2017}.
Directional TPCs have also been proposed for neutrino physics~\cite{Lisotti:2024fco}, and for neutrinoless double-beta decay~\cite{NEXT:2012rto}.

To achieve a PMT-only  3D reconstruction, we develop a reconstruction algorithm based on probabilistic graphical models, namely Bayesian Networks (BN)~\cite{pearlbook, jensenbook, DAgostini:2003syq, kollerbook}, that infers the $(X,Y)$ position of ionization tracks on the GEM plane from PMT signals, and estimates the light emitted during the amplification process, thus reconstructing the particle’s energy. This information is combined with the $\Delta Z$ component extracted from the analysis of PMT waveforms, particularly their time profile. Once the 3D reconstruction from PMTs is obtained, it can be matched with the camera image, which provides superior $(X,Y)$ spatial resolution and an independent energy measurement. This combination enables precise 3D reconstruction of the ionizing event and represents a promising path to further enhance the energy resolution, which will be quantitatively demonstrated in a forthcoming paper.
This methodology marks an important advancement for directional dark matter detection, enabling precise identification of nuclear recoil tracks with improved spatial resolution.

\section{Detector description}\label{sec:detector}

\begin{figure}
    \centering
    \includegraphics[width=\linewidth]{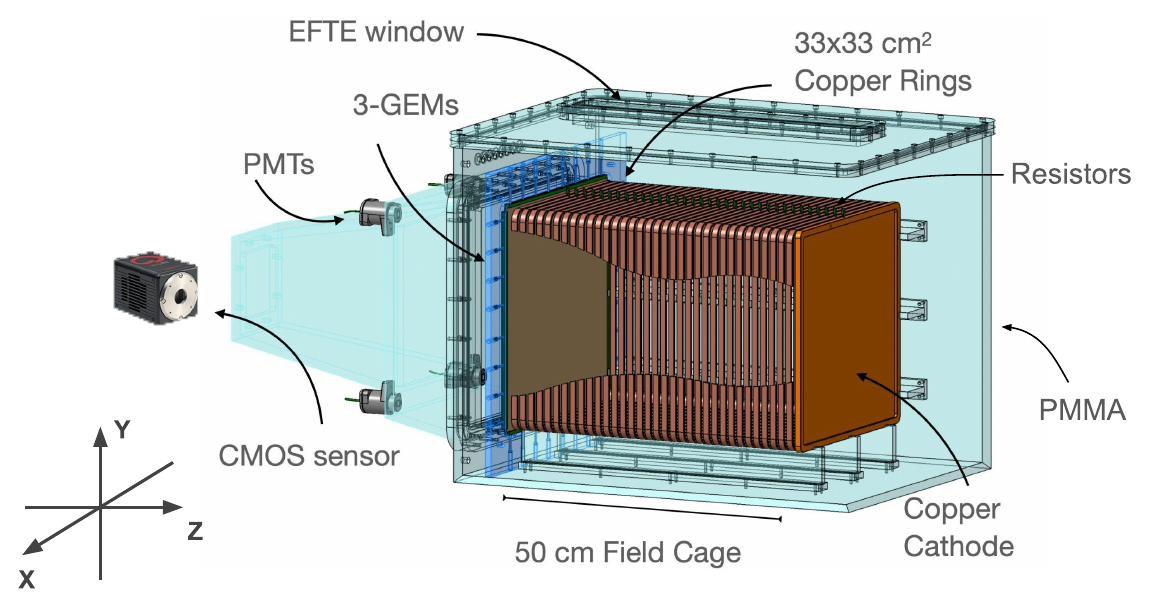}
    \caption{Schematic view of the LIME detector. The He:CF$_4$ (60:40) gas mixture is contained in a PMMA vessel housing a copper field cage. Ionization electrons drift from the cathode (right) toward the amplification region (left), where a triple-GEM structure produces charge multiplication and scintillation light. This light is collected by a centrally aligned APS-sCMOS camera and four PMTs located above the GEM plane, on the optical readout side.}
    \label{fig:LIME_setup}
\end{figure}

\subsection{The LIME prototype}

\begin{figure}
    \centering
    \includegraphics[width=0.5\textwidth]{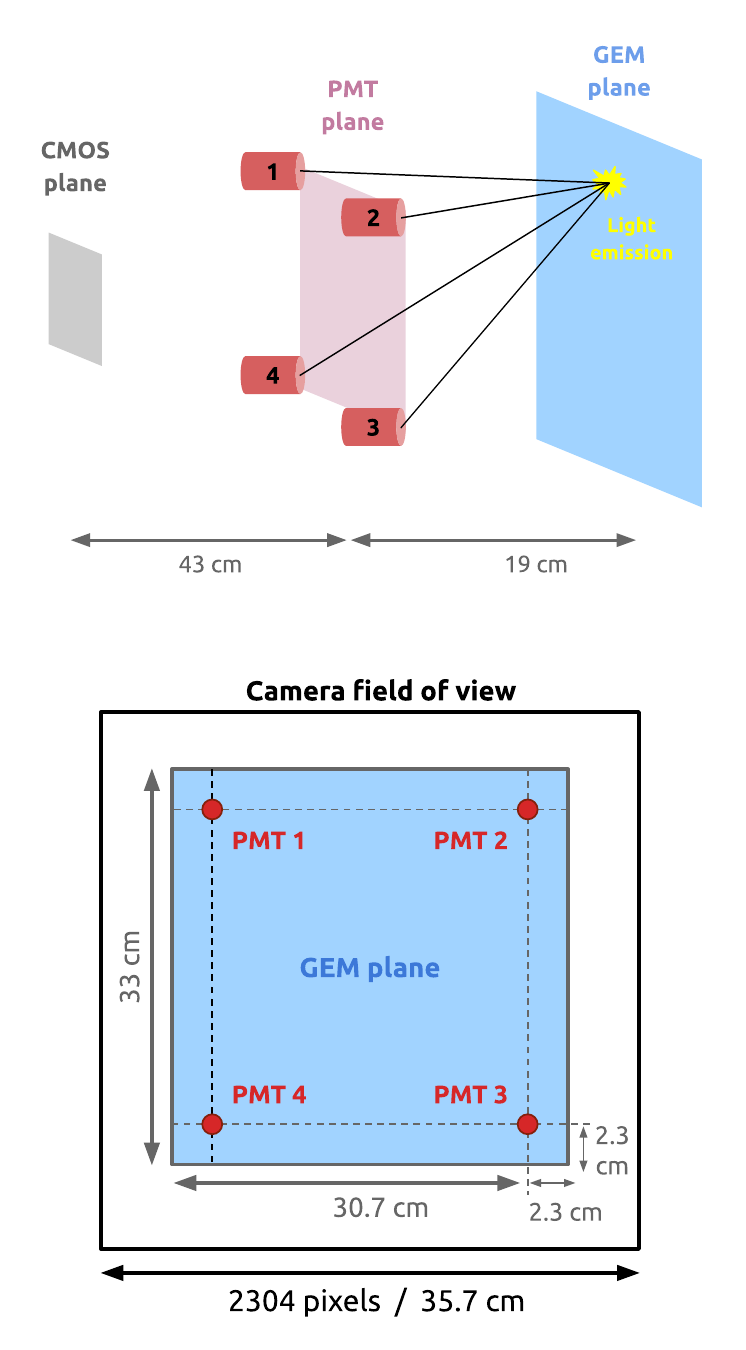}
    \caption{Relative disposition of the sensors with respect to the GEM plane, where light is emitted. Top: side view showing the field cage and the vertical distances between the PMTs and the GEMs. Bottom: front view, showing the camera position (centered) and the four PMTs (at the corners).}
    \label{fig:pmt_cam_fov}
\end{figure}

The Long Imaging ModulE (LIME)~\cite{LIME}, shown in Fig.~\ref{fig:LIME_setup}, is a prototype for the future CYGNO experiment. It consists of a box-shaped TPC with a drift distance of 50~cm and a transverse area of 33 $\times$ 33 \,cm$^2$. The He:CF$_4$ gas mixture is enclosed in a 10~mm-thick PMMA box, surrounded by a field cage composed of 34 copper rings with a cross-section of 330\,mm $\times$ 330\,mm, each 10\,mm thick and spaced 4\,mm apart. The field cage is bounded by a 0.5\,mm-thick copper cathode on one side and a triple-GEM stack on the other. Each GEM has holes of 50\,$\mu$m with a pitch of 140\,$\mu$m and is separated by a 2\,mm gap. A conical black PMMA structure is mounted on the side of the GEM stack to house a Hamamatsu ORCA-Fusion APS-sCMOS camera and four Hamamatsu R7378A PMTs, each with a 25.4\,mm diameter and a quantum efficiency of about 25\% in the 300–500\,nm wavelength range. The CF$_4$ emits light in two broad continua, peaked around 290\,nm and 620\,nm ~\cite{FRAGA200388_rydberg_state}, with the PMT sensitivity matching the UV component. The camera is equipped with a Schneider Xenon 0.95/25-0037 lens, featuring a focal length of 25.6\,mm and an aperture ratio (f-number) of 0.95. It is positioned centrally in front of the GEM plane at a distance of 62.3\,cm, while the four PMTs are located at the corners of a square plane parallel to the GEMs, 19\,cm away from them. In this configuration, the camera's field of view (FOV) covers an area of 35.7\,cm $\times$ 35.7\,cm, corresponding to a pixel granularity of 155\,$\mu$m. Side and front views of the geometric arrangement between the camera, PMTs, and the GEM plane are shown in Figure~\ref{fig:pmt_cam_fov}.

When a particle interacts within the LIME detector, it ionizes the gas mixture, and the resulting electrons drift toward the anode with a mean velocity of about $v = 5.5$ cm/$\mu$s under an electric field of 0.8 kV/cm, as estimated with Garfield simulations~\cite{Veenhof:1998tt, Baracchini:2020phn}. Upon reaching the GEM stage, the electrons undergo avalanche multiplication within the GEM holes~\cite{ELY}, where the local electric field reaches up to 40 kV/cm, also simulated with Garfield. During this process, scintillation light is emitted primarily by the CF$_4$ component of the gas. Part of the resulting photons are detected by the PMTs, enabling time-based reconstruction of the particle’s trajectory orthogonal to the GEM plane, and by the camera, which records the profile in the GEM plane.
\begin{figure*}
    \centering
    \begin{subfigure}{0.49\textwidth}
        \centering
        \includegraphics[width=\textwidth]{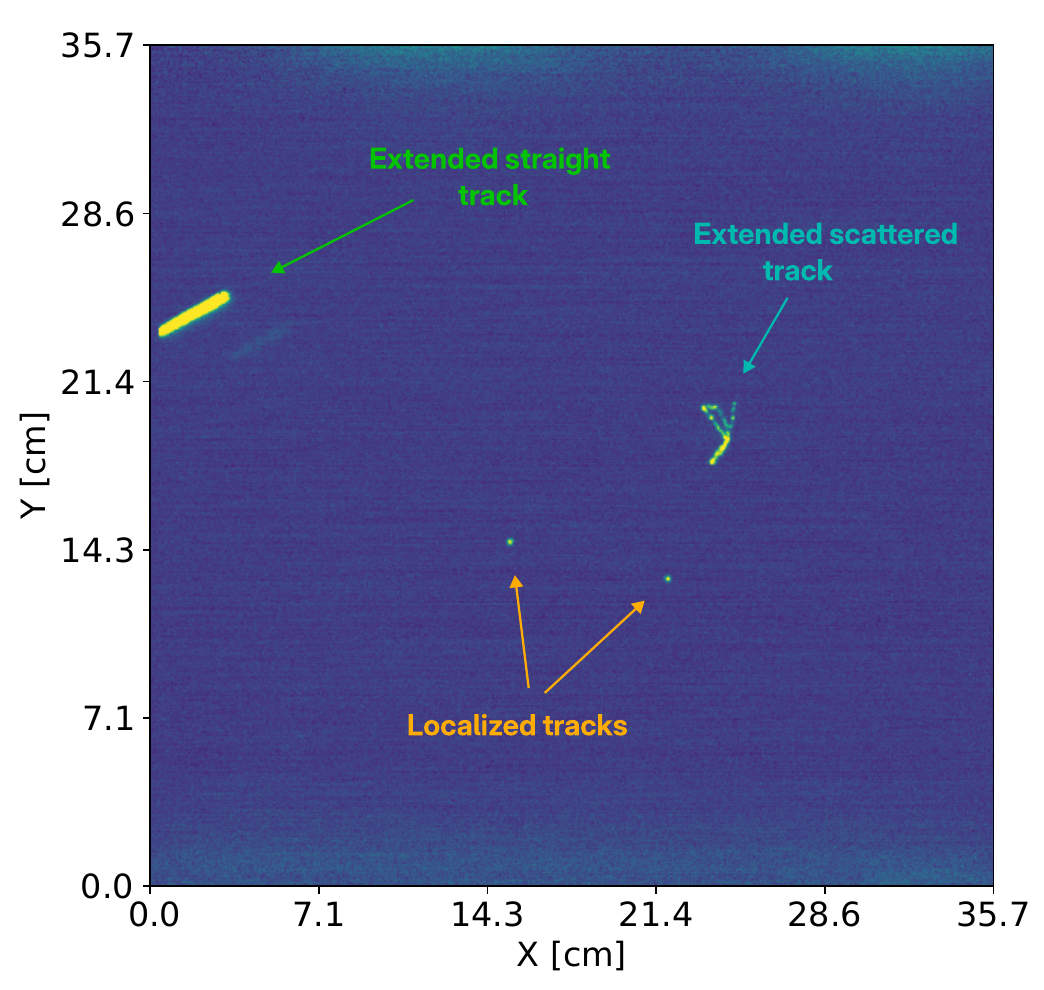}
        \caption{}
        \label{fig:LIME_evt_cmos}
    \end{subfigure}\hfill
    \begin{subfigure}{0.5\textwidth}
        \centering
        \raisebox{2mm}{\includegraphics[width=\textwidth]{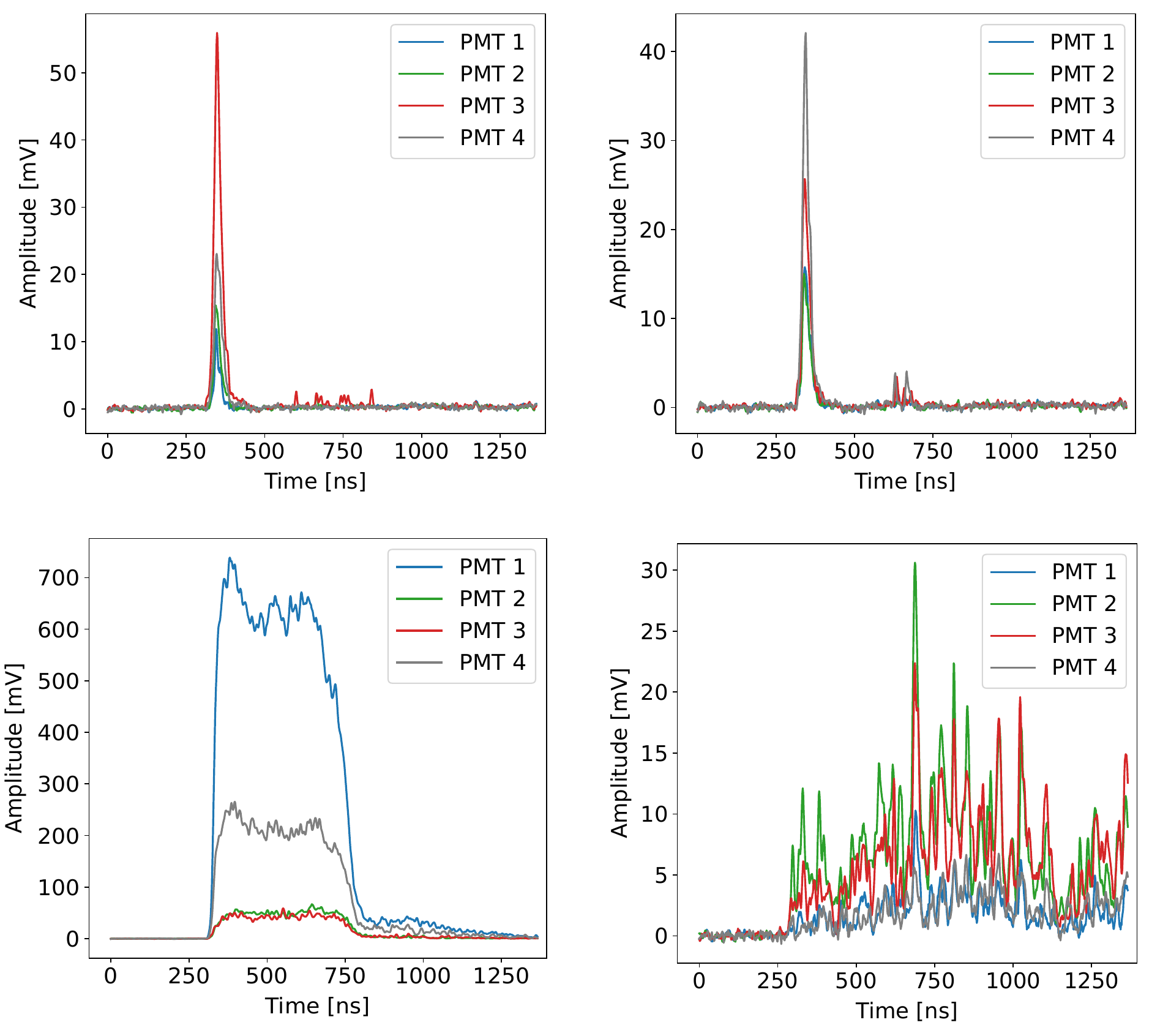}}
        \caption{}
        \label{fig:LIME_evt_pmts}
    \end{subfigure}
   \caption{Example of an event recorded with the LIME's optical readout, illustrating (\subref{fig:LIME_evt_cmos}) the image acquired by the APS-sCMOS camera during a 300\,ms exposure with four distinct tracks: two localized clusters; one extended straight ionization trail; and a curly scattered track (electron recoil). Figure (\subref{fig:LIME_evt_pmts}) shows the PMT signals (inverted for clarity) recorded within the same acquisition window, each associated to one of the ionization in the picture.}
    \label{fig:event}
\end{figure*}

An event in LIME consists of a combination of a camera image and a set of PMT signals. The data acquisition system (DAQ) operates as follows: the camera runs in trigger-less mode with a 300\,ms exposure (the minimum value allowed in the low noise configuration), while PMT signals are continuously monitored. These signals pass through a leading-edge discriminator, and a trigger is issued if at least two PMTs exceed a predefined voltage threshold within the same time window. When a trigger occurs, the image is saved along with the corresponding PMT signals recorded over an 1.40\,$\mu$s time window. In cases where multiple PMT triggers occur within the same camera exposure, several sets of PMT signals are associated with a single image. Figure~\ref{fig:event} shows an example of such an event recorded in LIME, where multiple tracks are visible in the image -- two localized clusters, one extended track induced by an alpha particle, and one curly electron recoil-like event -- along with the corresponding PMT signals generated by these ionization processes. In these situations, the Bayesian inference algorithm can be employed to associate each set of waveforms with its corresponding track observed in the image. To analyze such events, the CYGNO collaboration developed a reconstruction algorithm~\cite{Baracchini_2021} that identifies light clusters in the APS images and reconstructs their physical properties, including shape, light intensity, and direction on the GEM plane. The analysis of PMT signals has been developed in parallel with the present work and will be discussed in detail in a forthcoming publication.

\subsection{PMTs signals}\label{sec:PMT-signal}

\begin{figure}
    \centering
    \includegraphics[width=0.8\linewidth]{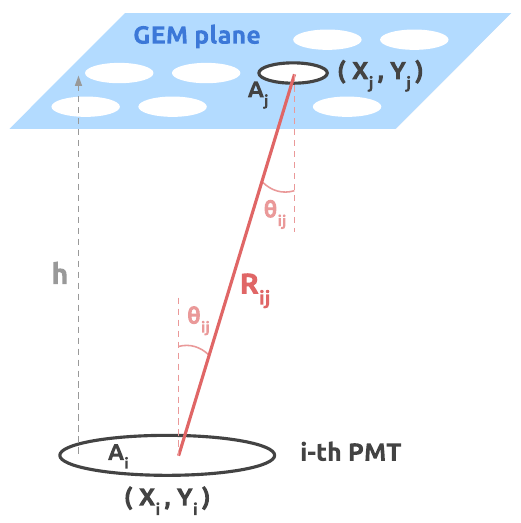}
   \caption{Schematic representation of the illumination of the $i$-th PMT by the radiating source with coordinates $(X_j,Y_j)$ on the GEMs. The distance between the centers of the two surfaces is denoted by $R_{ij}$, and the angle with respect to the $z$-axis is $\theta_{ij}$.}
    \label{fig:optical-scheme}
\end{figure}

The relationship between the light per unit surface per steradian $L_j$ emitted from position $(X_j,Y_j)$ of the GEMs and the corresponding light $L_{ij}$ collected by the $i$-th PMT is modeled under the assumption of Lambertian emission, i.e. radiation from a perfectly diffuse source with angle-independent radiance~\cite{optics}. In this case, the light received at a distance $R_{ij}$, as shown in Fig.~\ref{fig:optical-scheme}, is
\begin{equation}
L_{ij} = \frac{L_j A_j A_i \cos^2{\theta_{ij}}}{R_{ij}^2},
\end{equation}
where $A_j$ and $A_i$ denote the source and PMT areas. With $\cos \theta_{ij}=h/R_{ij}$, where $h=19$\,cm is the fixed PMT–GEM separation, the expression reduces to
\begin{equation}
\label{eq:fourth-power-law}
L_{ij} \propto \frac{L_j}{R_{ij}^4},
\end{equation}
the cosine-fourth-power law of illumination.  The PMT angular response was neglected, as it is modest at the viewing angles of interest~\cite{pmt_handbook}.
The signal in the $i$-th PMT is a voltage $V_{ij}(t)$ (Fig.~\ref{fig:LIME_evt_pmts}). The corresponding charge $Q_{ij}$, proportional to $L_{ij}$, is
\begin{equation}
\label{eq:li_definition}
 Q_{ij} =  \frac{1}{\mathcal{R}}\int_{\Delta t} V_{ij}(t)\,dt\, \propto L_{ij}
\end{equation}
where $\mathcal{R}=50\,\mathrm{\Omega}$ the termination impedance, and $\Delta t$ the integration window. Combining Eqs.~\ref{eq:fourth-power-law} and \ref{eq:li_definition} yields
\begin{equation}
\label{eq:charge-light}
Q_{ij}= C_i \frac{L_j}{R_{ij}^{\alpha}},
\end{equation}
with $\alpha=4$ and $C_i$ calibration constants which include the relative gain and efficiency of $i$-th PMT as well as a global calibration factor (absolute scale) and the proportionality factors omitted in Eqs.~\ref{eq:fourth-power-law}, \ref{eq:li_definition}. Their dimensions are $[C_i] = [\mathrm{L}]^4[\mathrm{Q}]/[L_j]$, with $[L_j]$, $[\mathrm{Q}]$, and $[\mathrm{L}]$ the dimensions of radiance, charge, and length, respectively.

The PMT high voltages have been pre-adjusted with LED bench tests in order to equalize the individual gains. 
Subsequently, dataset of nearly mono-energetic, localized events is used to characterize the PMT response. 
Such events are obtained by exposing the detector to a $^{55}$Fe radioactive source, 
which emits $K_\alpha$ and $K_\beta$ X-rays from $^{55}$Mn. 
Due to the limited energy resolution, these lines merge into an effectively monochromatic peak at 5.9~keV. 
A sample of $^{55}$Fe events is selected following the procedure described in Refs.~\cite{LIME,bib:fe55}.  
The PMT waveforms produced by these electronic recoils typically show a single peak 
(Fig.~\ref{fig:event}). 
The integrated charge is computed within a 60~ns window centered on the peak. 
Events are further required to exhibit a single light spot on the GEM plane, 
as reconstructed by the optical camera, with well-defined positions $(X_j,Y_j)$.  

Using this dataset and the $(X_j,Y_j)$ positions as reconstructed by the camera, we validate the cosine-fourth-power law of illumination by fitting the exponent $\alpha$ in the generalized form $L_{ij} \propto L_j/R_{ij}^\alpha$, obtaining $\alpha = 4.0$, consistent with the model within less than 10\%.

\begin{figure}
    \centering
    \includegraphics[width=\linewidth]{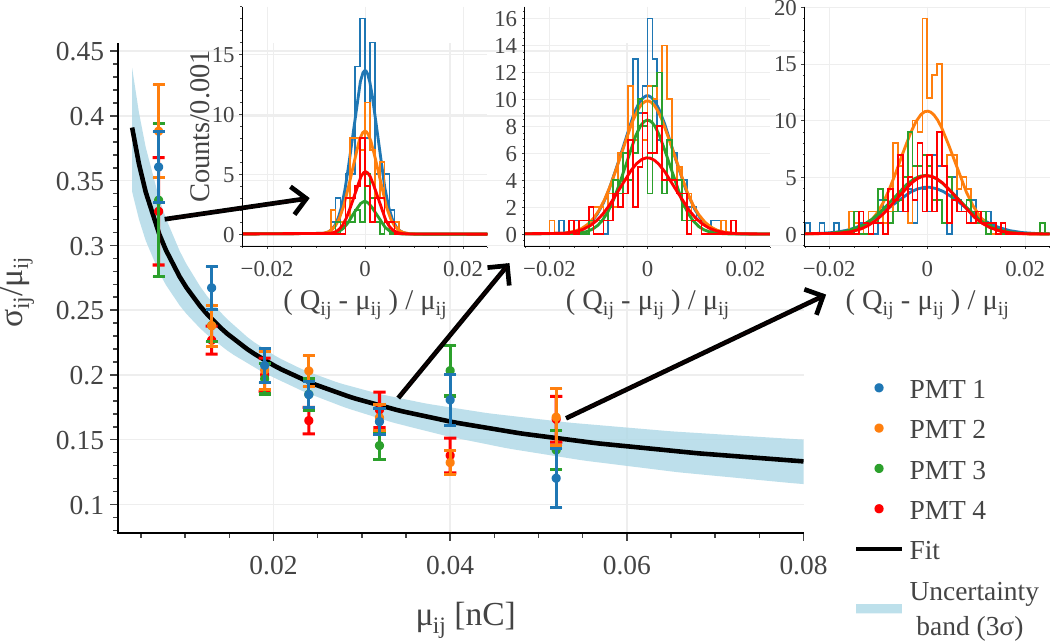}
    \caption{Relative standard deviation, for each PMT, of the measured charge as a function of the expected value $\mu_{ij}$ from Eq.~\ref{eq:charge-light}. The expected value spans over the dynamic range because of a $^{55}$Fe mono-energetic source occurring at different positions on the GEM plane. The colored points are the measured data, the black curve is the fit of the model described in Eq.~\ref{eq:dispersion} for all the PMTs and the blue band represents the $3\sigma$ uncertainty range of the inference process. The insets display the relative charge dispersion for the three points indicated by the arrows with the relative Gaussian model superimposed.}
    \label{fig:dispersion}
\end{figure}

The PMT response is then characterized by comparing the measured charge $Q_{ij}$ 
with its expected value $\mu_{ij}$ defined in Eq.~\ref{eq:charge-light}. 
Figure~\ref{fig:dispersion} shows, for each PMT, the relative standard deviation 
$\sigma_{ij}/\mu_{ij}$ as a function of $\mu_{ij}$. 
The dynamic range is sampled by interactions of 5.9~keV X-rays at different positions 
on the GEM plane. 
The insets display the charge dispersion for three selected points; 
in all cases, the distributions are well described by a Gaussian model.  
The general model adopted to describe the charge resolution is:
\begin{equation}
\frac{\sigma_{ij}}{\mu_{ij}}= \frac{A}{\sqrt{\mu_{ij}}} + B
\label{eq:dispersion} 
\end{equation}
where $A$ is the stochastic  and $B$ the constant term accounting for systematic effects in the response model. In principle, a pair $(A_i, B_i)$ would be required to describe each PMT independently. However, given the comparable energy resolution observed across the photo-sensors, a single response function was found to adequately describe all of them. The fitted parameter values are:
\begin{align}
    A=& (0.021 \pm 0.002)\,\sqrt{\text{nC}} \nonumber\\
    B=& (0.06 \pm 0.01)\nonumber \\
    \rho_{AB}=&-97\%
\label{eq:inferredAB}
\end{align}
We also tested an extended three-parameter fit including an additional term proportional to ${\mu^{-1}_{ij}}$ to account for charge-independent effects such as PMT electronic noise, which dominate at very low signals. However, no appreciable improvement is observed in the residuals, which remain equally well described by the two-parameter model. Since our study is limited to the same charge range shown in Fig.~\ref{fig:dispersion}, the two-parameter description is both simple and effective, though it should not be regarded as a general measurement of the detector resolution outside this range.
\section{Application of Bayesian Networks for PMT-Based 3D Reconstruction} 

In Bayesian inference, Bayes' theorem is used to update the probability of a model or a set of parameters as new evidence or information becomes available, in the form of experimental observations. 
Denoting with $\{x\}$ the data, and with $\boldsymbol{\theta}$ the parameters describing the experimental conditions or the theoretical assumptions, the joint posterior probability $p(\boldsymbol{\theta}|\{x\})$ derives from the product of the direct probability $p(\{x\}|\boldsymbol{\theta})$, the prior probability $\pi(\boldsymbol{\theta})$ and the data probability $p(\{x\})$  as:
\begin{equation}\label{eq:posterior1}
    p(\boldsymbol{\theta}|\{x\})=\frac{p(\{x\}|\boldsymbol{\theta}) \cdot \pi(\boldsymbol{\theta})}{p(\{x\})} . 
\end{equation}
When the direct probability $p(\{x\}|\boldsymbol{\theta})$ is regarded as a function of the parameters $\boldsymbol{\theta}$ for some given data $\{x\}$ it is referred to as the likelihood function $\mathcal{L}(\boldsymbol{\theta}; \{x\})$.

In the case of the PMT-based reconstruction in the LIME prototype, the likelihood used in the inference process of Eq.\,~\ref{eq:posterior1} is defined as:
\begin{equation}\label{eq:likelihood1}
    \begin{split}
        \mathcal{L}(\boldsymbol{\theta}, A, B; \{Q_{ij}\}) &= \prod_{j=1}^{N} \prod_{i=1}^{4} \mathcal{N}(Q_{ij} \mid \mu_{ij}( \boldsymbol{\theta}), \sigma_{ij}( \boldsymbol{\theta}, A, B)) \\
    \end{split}
\end{equation}
where $j$ runs over all light-emitting sources, $i$ denotes the PMT index, and $Q_{ij}$ is the charge measured by the $i$-th PMT for the $j$-th emission point. All parameters describing the model and the experiment are collected in $\boldsymbol{\theta}$: the geometry $(X_i, Y_i, h)$, which is considered known and kept fixed; the global exponent $\alpha$, also fixed; the calibration parameters $(C_i)$, with priors taken from dedicated measurements; and the source variables $(X_j, Y_j, L_j)$. The observed charges $Q_{ij}$ are assumed to be independently normally distributed around the expected values $\mu_{ij}( \boldsymbol{\theta})$ from Eq.~\ref{eq:charge-light}, with standard deviation $\sigma_{ij}( \boldsymbol{\theta},A, B)$ described by Eq.~\ref{eq:dispersion}. The parameters $A$ and $B$ of the resolution function are passed to the fit as nuisance parameters with the bivariate normal prior specified in Eq.~\ref{eq:inferredAB}. 
%

\begin{figure}
    \centering
    \includegraphics[width = 0.35\textwidth]{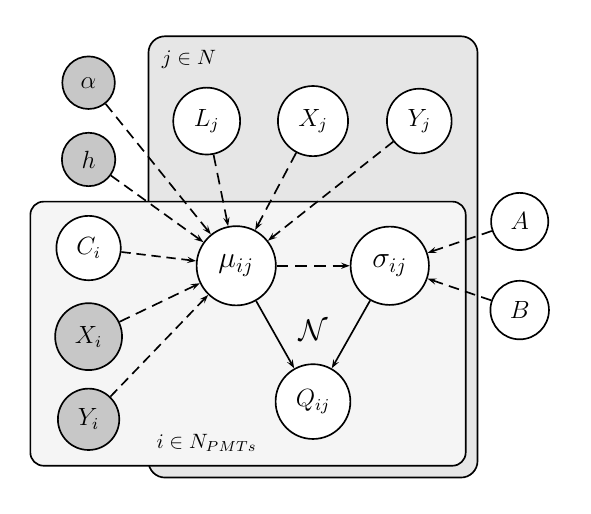}
    \caption{Bayesian network adopted in the PMT reconstruction. Each event $j \in N$ produces a light yield $L_j$ at position $(X_j, Y_j)$, while each PMT $i \in N_{PMTs}$ is characterized by a calibration coefficient $C_i$ and position $(X_i, Y_i)$. The expected mean signal $\mu_{ij}$ depends on these quantities, on global parameters $(\alpha, h)$, while $\sigma_{ij}$ depends also on nuisance terms $(A, B)$. The observed charge $Q_{ij}$ is modeled as a Gaussian distribution $\mathcal{N}$ with mean $\mu_{ij}$ and width $\sigma_{ij}$. Plates indicate repetition over events ($j$) and over PMTs ($i$); solid arrows indicate probabilistic links, dashed arrows deterministic ones. Primordial nodes are either fixed (grey) or assigned a prior (white).
   }
    \label{fig:BN_PMT}
\end{figure}

Fig.~\ref{fig:BN_PMT} shows the graphical representation of the likelihood through a Bayesian network. 
Plates indicate repetition over events ($j$) and over PMTs ($i$); solid arrows indicate probabilistic links, dashed arrows deterministic ones; primordial nodes are either fixed (grey) or assigned a prior (white). Inference in this graphical model proceeds in both directions: once a node is observed, the corresponding information propagates backward through the network. For instance, measuring the charges $Q_{ij}$ allows us to infer the underlying event variables $X_j$, $Y_j$, and $L_j$, so that measurements constrain not only the directly connected nodes but also their ancestors.

The Bayesian network formalism is particularly useful as it makes the probabilistic assumptions explicit and allows flexible and modular extension of the model. This approach differs significantly from classical reconstruction strategies commonly used in optical TPCs, such as centroid estimators, clusterization algorithms, or $\chi^2$-based fitting methods. While those techniques are computationally efficient, they often lack a consistent way to propagate uncertainties and may struggle with complex detector geometries or signal topologies. In contrast, the Bayesian network encodes the underlying physics and geometry explicitly, enabling the direct computation of posterior distributions for all parameters of interest. Despite the use of sampling methods, the relatively small number of parameters per event makes the inference tractable. Moreover, the modular structure of the Bayesian model allows for transparent extensions to more complex scenarios such as multiple tracks, energy-dependent emission models, or prior-informed inference in low-signal regimes.

Possible extensions in our model may also include additional uncertain parameters, such as the detector geometry $(X_i, Y_i, h)$ or the global exponent $\alpha$, which are currently treated as fixed: the geometry is precisely determined and its uncertainty is negligible, while for the exponent $\alpha$ we rely on the strong theoretical prior of the Lambertian model ($\alpha=4$), consistent with our data within $\sim$10\%.

\section{Code implementation}

The reconstruction algorithm is implemented  using the BAT software~\cite{ref:BAT,2009CoPhC.180.2197C}, a comprehensive package of C++ libraries tailored for Bayesian analysis. BAT has been widely adopted in the high energy physics community, with applications in numerous collaborations, including UTfit~\cite{Ciuchini:2000de} and HEPfit~\cite{deBlas:2021wap}, as well as in direct detection experiments such as DarkSide-50~\cite{DarkSide-50:2023fcw} and XENONnT~\cite{XENONCollaboration:2023dar}. The parameter inference is performed using the Metropolis-Hastings Markov Chain Monte Carlo (MCMC) algorithm. To ensure proper convergence of the MCMC chains, BAT includes a pre-run phase that automatically tunes the sampling parameters. During this phase, the step sizes and other internal variables are optimized to guide all chains toward the same region of the parameter space, achieving an optimal acceptance rate for the Metropolis-Hastings proposals.

\section{Fitting strategy and dataset}

The Bayesian network of Eq.~\ref{eq:likelihood1} and Fig.~\ref{fig:BN_PMT} can be implemented in three configurations:
(1) PMT calibration, where the fit runs on $n$ point-like events of known light $L_j$ and position $(X_j,Y_j)$ (thus $N=n$ in Eq.~\ref{eq:likelihood1}) to determine the calibration constants $C_i$ (Sec.~\ref{sec:PMT-calibration});
(2) reconstruction of spot-like events to infer $L_j$ and $(X_j,Y_j)$ (Sec.~\ref{sec:loc_track_rec});
(3) reconstruction of extended tracks to infer $L_j$ and $(X_j,Y_j)$ of their sub-segments (Sec.~\ref{sec:extended-track}).
For modes (2) and (3), the fit is performed with $N=1$ in Eq.~\ref{eq:likelihood1}.

\subsection{In-situ calibration of the PMT response}\label{sec:PMT-calibration} 
A first application of the model is the in-situ calibration of the PMT response, whose relevance lies in accounting not only for gain equalization but also for geometric and alignment effects, as well as material-dependent factors influencing light collection. The calibration consists in the determination of the coefficient $C_i$ for each PMT. 

This is achieved by analyzing events from a known position in the $(XY)$ plane with identical energy deposition $L$ and track topology. For this purpose, we select a dataset of $N=n$ events characterized by a single light cluster in the camera image and a single signal recorded by each of the four PMTs. These events are acquired during exposure to a $^{55}$Fe radioactive source similarly to what described in section~\ref{sec:PMT-signal}.

\begin{figure}
    \centering
    \includegraphics[width = \linewidth]{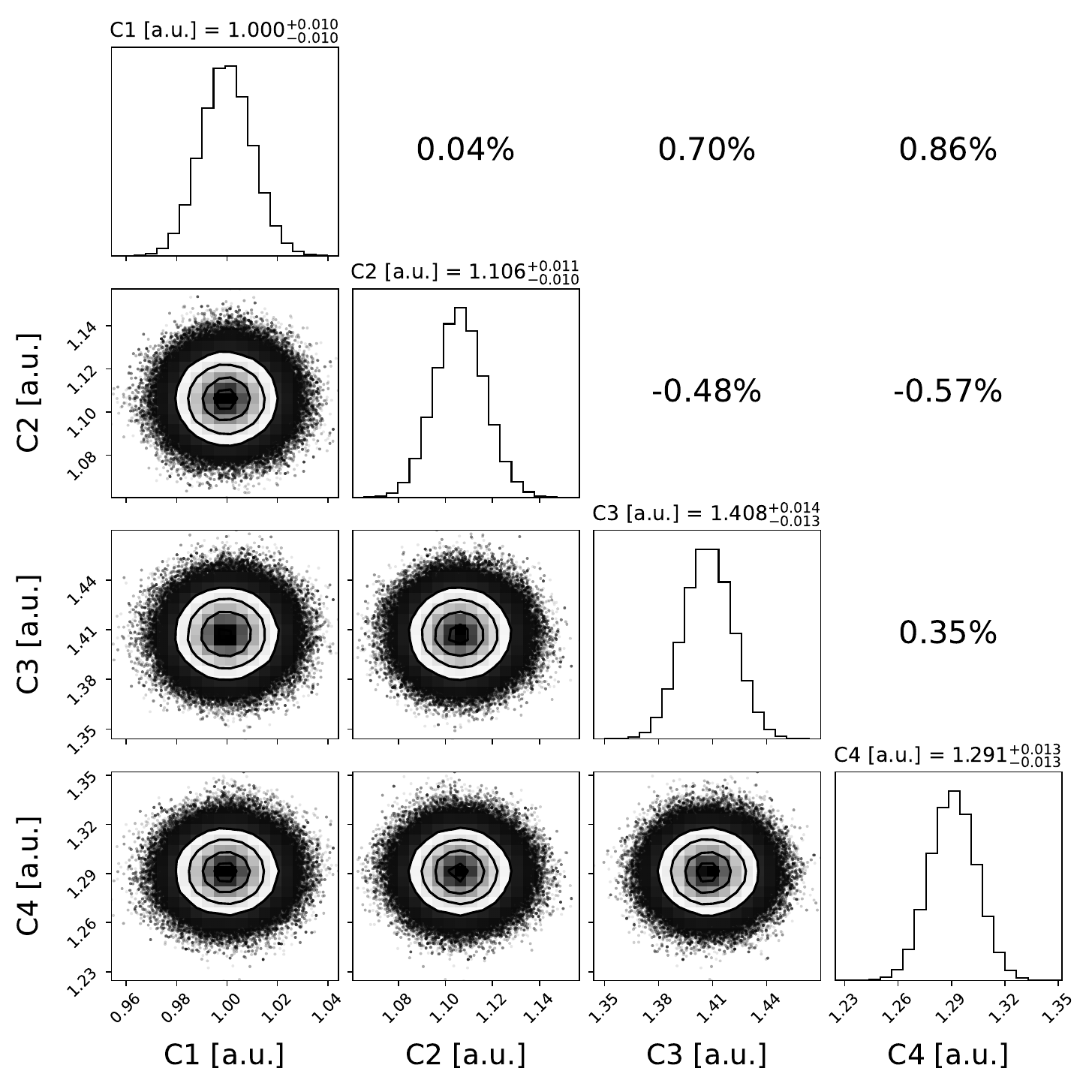}
    \caption{Corner plot of the posterior distributions obtained from the calibration algorithm, normalized to $C_1$. The diagonal panels show the 1D histograms of each PMT calibration parameter $C_i$, while the off-diagonal panels display the scatter plots of the corresponding parameter pairs, along with their correlation. The labels in the diagonal histograms report the 16th, 50th, and 84th percentiles of each distribution.}
    \label{fig:calib_posterior}
\end{figure}

An example of the inferred posterior distributions is shown in Fig.~\ref{fig:calib_posterior}, where the mean values are normalized to $C_1$. The 16th, 50th, and 84th percentiles are indicated for each parameter, together with the correlation coefficients displayed in the upper-right subplots.

The calibration fit is performed using 12 parallel MCMC chains, each consisting of 100.000 steps. A standard calibration procedure based on 669 events required 4 minutes and 46 seconds of user CPU time on a single core. All computations were performed on a machine equipped with an Intel(R) Xeon(R) E5-2620 CPU running at 2.00 GHz.
    
\subsection{Reconstruction of localized tracks}\label{sec:loc_track_rec}
    
Once the calibration constants are fixed (e.g., using the values obtained in the previous step), the Bayesian framework can be employed to reconstruct the position and intensity of localized tracks. For each set of four PMT signals, the parameters are inferred using Eq.~\ref{eq:likelihood1}, with $N = 1$ source. Flat priors are assigned to $X$ and $Y$, constrained within the GEM plane (33 $\times$ 33\,cm$^2$), while the prior on light intensity is flat over positive values below a predefined upper bound.
For this task we use events collected during exposure to $^{55}$Fe radioactive source.

\begin{figure}
    \centering
    \includegraphics[width=\linewidth]{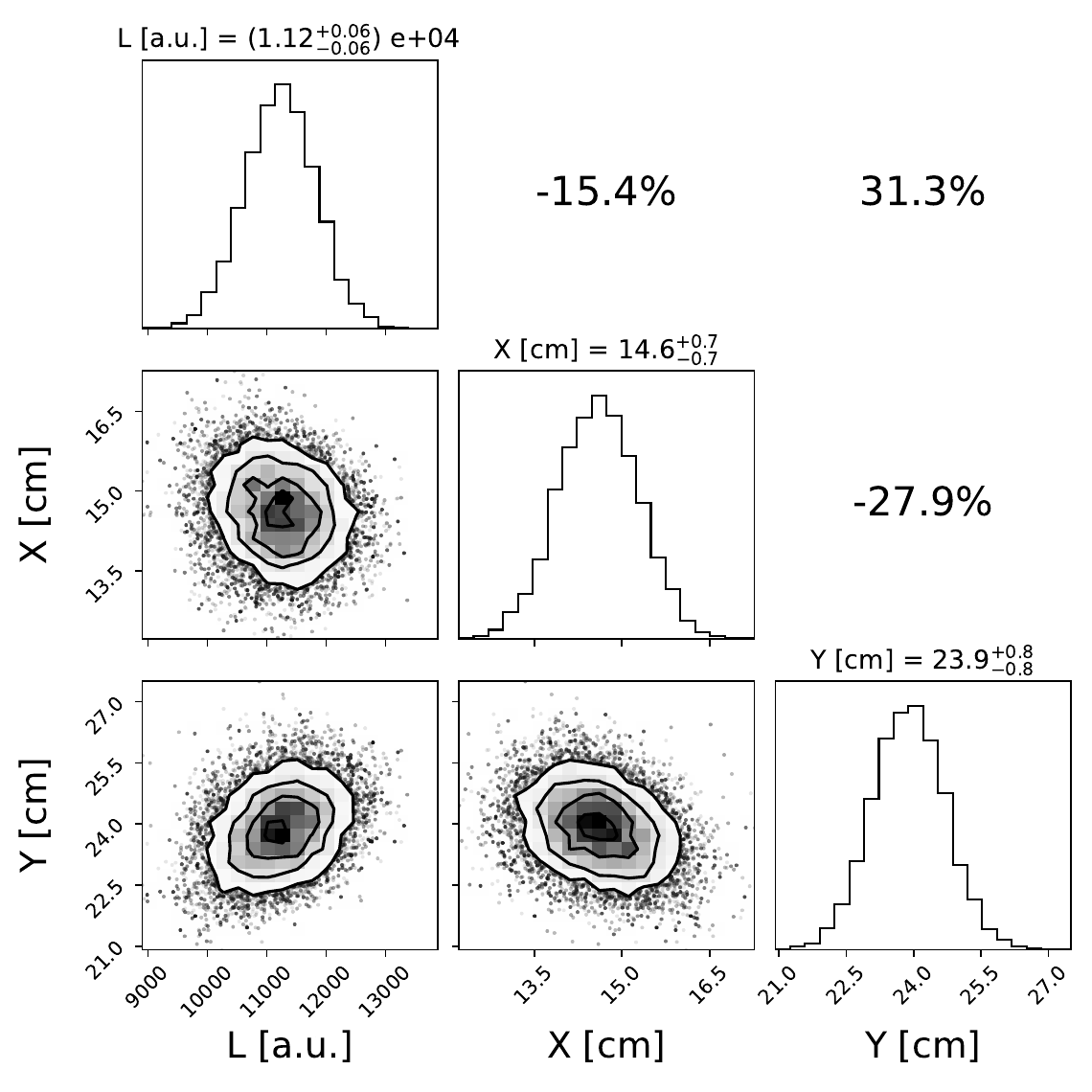}
    \caption{Corner plot of the posterior distributions obtained from the reconstruction algorithm applied to localized tracks. The diagonal panels show the 1D histograms of the $X$, $Y$, and $L$ parameters, while the off-diagonal panels display the corresponding scatter plots and their correlations. Each histogram is labeled with the 16th, 50th, and 84th percentiles of the respective distribution.}
    \label{fig:ass_corner}
\end{figure}

As in the calibration step, the integral of the PMT signal is computed within a 60\,ns time window centered on the main waveform peak. A representative example of the resulting posterior distributions for a single event is shown in Fig.~\ref{fig:ass_corner}. For this specific event, some correlations are visible among $X$, $Y$, and $L$. These correlations arise from the structure of the likelihood and the event topology in the $(XY)$ plane, and vary on an event-by-event basis. We verified that, when averaging over all events, the correlations vanish.

\begin{figure}
    \centering
    \includegraphics[width=\linewidth]{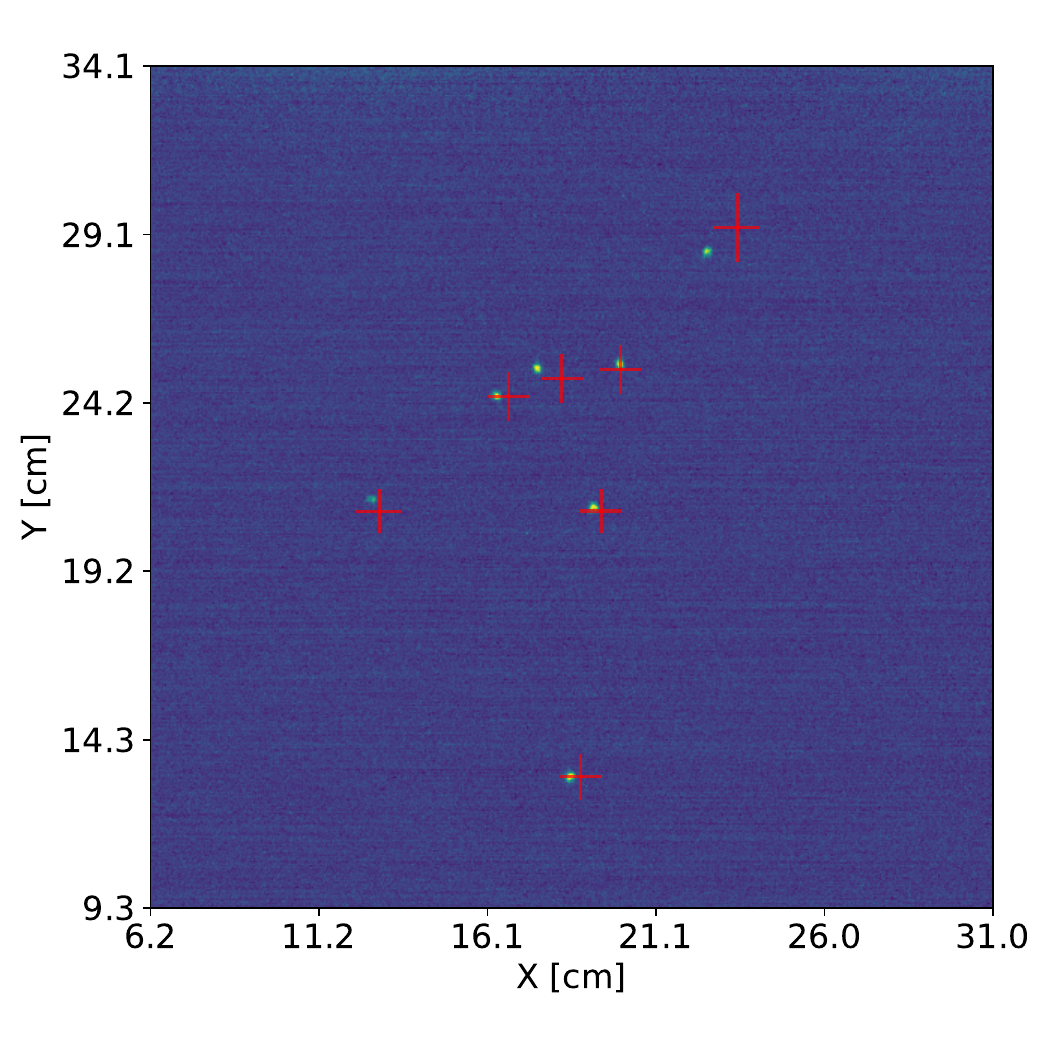}
   \caption{Reconstructed $(X,Y)$ positions obtained through the Bayesian fitting procedure (shown as \textit{red crosses}) overlaid on the camera image. The size of each cross represents the uncertainty of the fit in both dimensions. The yellow dots visible in the image correspond to the highly localized electron recoils induced by the $^{55}$Fe radioactive source.} 
    \label{fig:bat_on_cmos_calib}
\end{figure}

An example of the $(X,Y)$ reconstruction from PMT data is shown in Fig.~\ref{fig:bat_on_cmos_calib}, where the inferred positions are overlaid on the camera image. The reconstructed coordinates are in good agreement with the positions of the electron recoils induced by the $^{55}$Fe source, clearly visible in the image.

\begin{figure*}
    \centering
    \begin{subfigure}{0.33\textwidth}
        \centering
        \raisebox{0mm}{\includegraphics[width=\textwidth]{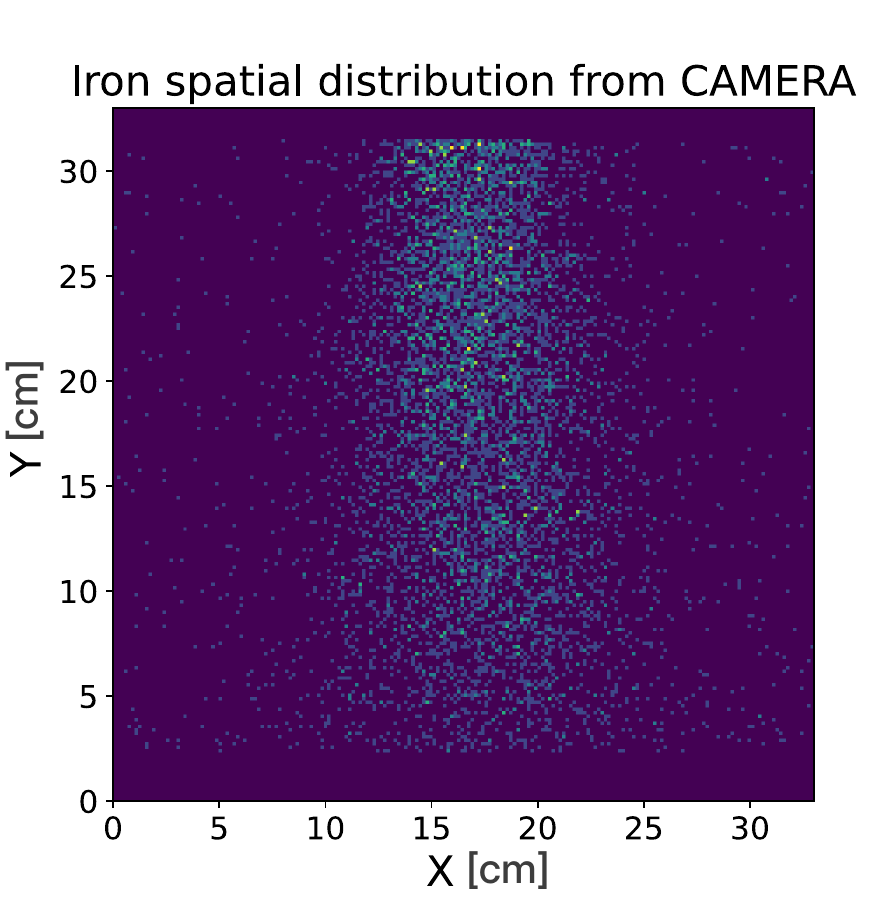}}
        \caption{}
        \label{fig:fe_distribution_1}
    \end{subfigure}\hfill
    \begin{subfigure}{0.33\textwidth}
        \centering
        \includegraphics[width=\textwidth]{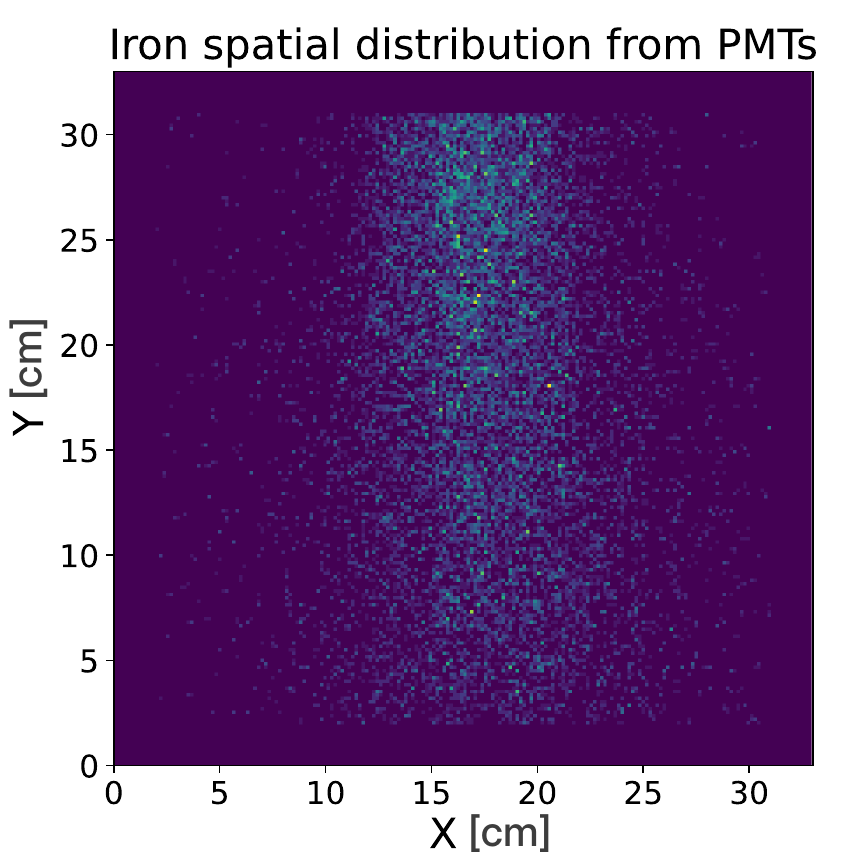}
        \caption{}
        \label{fig:fe_distribution_2}
    \end{subfigure}\hfill
    \begin{subfigure}{0.33\textwidth}
        \centering
        \includegraphics[width=\textwidth]{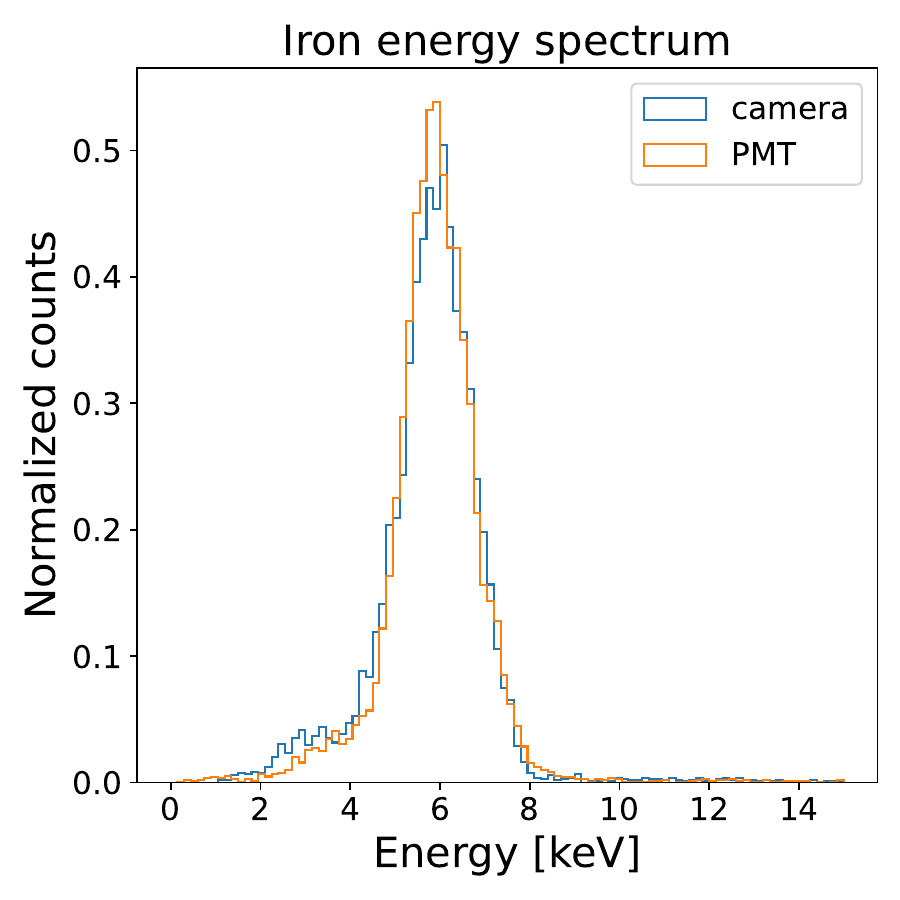}
        \caption{}
        \label{fig:fe_distribution_3}
    \end{subfigure}
   \caption{Planar distribution of the tracks in the camera field of view reconstructed using (\subref{fig:fe_distribution_1}) the APS-sCMOS analysis, and (\subref{fig:fe_distribution_2}) the PMT-based Bayesian algorithm, using a dataset acquired in with a $^{55}$Fe radioactive source positioned above the detector. (\subref{fig:fe_distribution_3}) Reconstructed energy spectrum with both analysis.}
    \label{fig:fe_distribution}
\end{figure*}

Figure~\ref{fig:fe_distribution_1} shows the spatial distribution of the spots reconstructed with the camera-based algorithm and figure~\ref{fig:fe_distribution_2} the PMT-based algorithm. The distribution appears narrower along the x-axis due to the presence of a collimator in front of the source. As a result, the spot density is higher along X, which -- as will be discussed later -- leads to better spatial resolution in that direction compared to Y. In addition, figure~\ref{fig:fe_distribution_3} shows that the reconstructed energies from the two sub-detectors are in very good agreement. For the CMOS, the energy estimation is obtained from the total photon counts in the pixels constituting the tracks, while for the PMT it comes directly from the BAT-fit light output variable $(L)$. A single-point calibration is applied by associating the main peak in both sensors distributions to the $^{55}$Fe peak of 5.9\,keV. The observed non-gaussian tails are due to poorly reconstructed tracks and background events. This result demonstrates that the limiting factor for the energy resolution is the number of photons produced at the end of the charge amplification process, rather than the readout method.

\begin{figure}[ht]
    \centering
    \includegraphics[width = \linewidth]{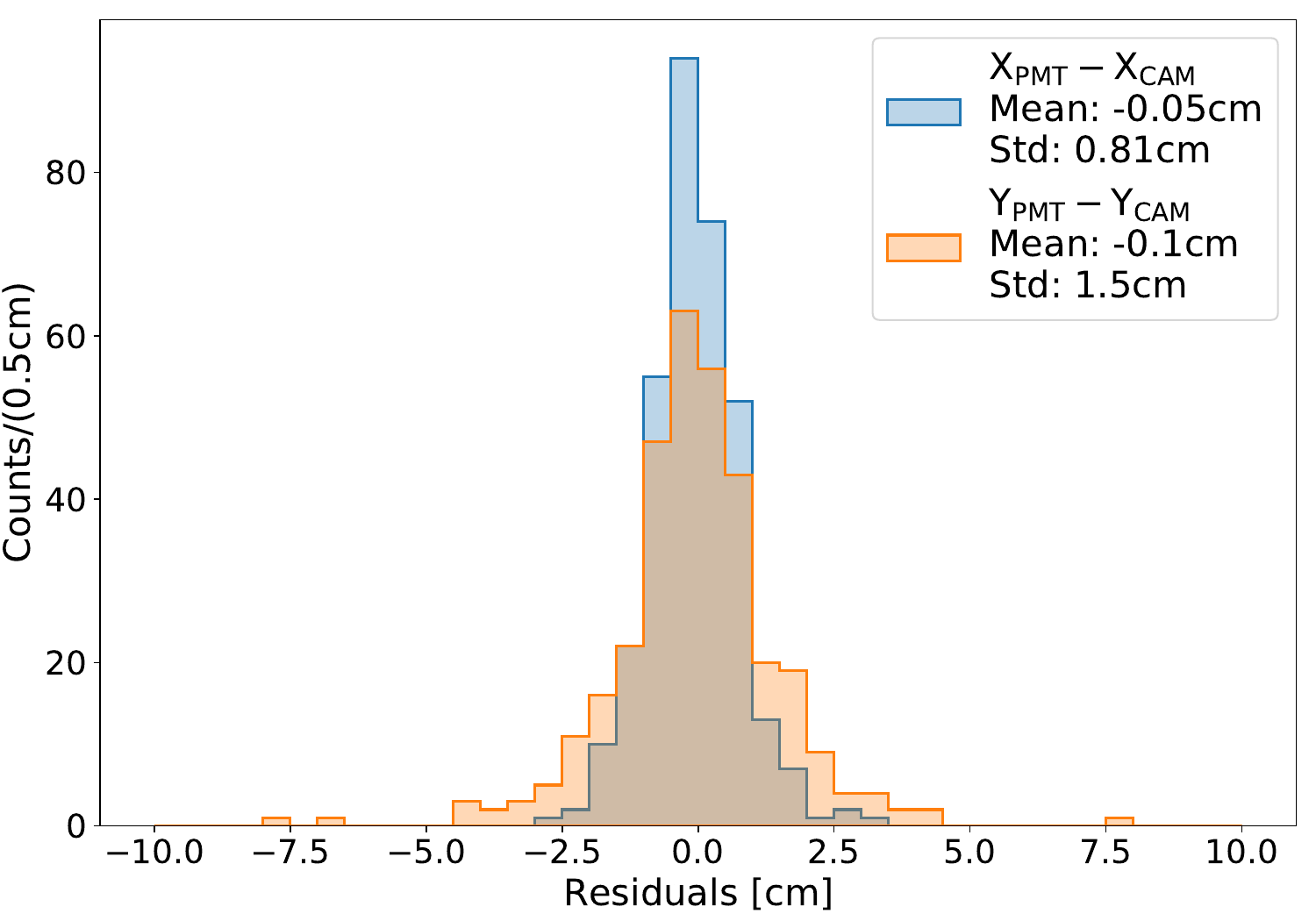}
    \caption{Distribution of the residuals $\Delta X$ and $\Delta Y$ between the PMT-based and camera-based track reconstructions for localized events. The dataset includes only events featuring a single localized track in the image and a single PMT waveform, allowing for a direct match between the two sensors' information.}
    \label{fig:golden_association}
\end{figure}

Events featuring a single visible localized track and a single PMT trigger are used to evaluate the reconstruction accuracy. Figure~\ref{fig:golden_association} shows the distributions of the residuals $\Delta X$ and $\Delta Y$ between the PMT-based and camera-based reconstructions of the $X$ and $Y$ coordinates, respectively. The mean values and standard deviations are
\begin{align}
\Delta X & = (-0.05 \pm 0.81)~\text{cm}, \\
\Delta Y & = (-0.1 \pm 1.5)~\text{cm}.
\end{align}
The quoted uncertainties represent the $(X,Y)$ spatial resolution achievable with the PMT-only approach. The difference between the two coordinates reflects the $X$-axis collimation of the source, which results in a different spatial spread of the events used for this study (see Figs.~\ref{fig:fe_distribution_1} and \ref{fig:fe_distribution_2}).

In addition, a toy Monte Carlo simulation was developed to validate the internal consistency of the reconstruction algorithm. By generating synthetic spot-like emissions with known positions and intensities, and applying the Bayesian reconstruction framework, we confirmed that the algorithm accurately recovers both the source positions and the emitted light within the expected uncertainties, further supporting the reliability of the PMT-based approach.

Each single-event fit is performed using six parallel MCMC chains, each consisting of 10.000 steps. To assess the reconstruction performance, a total of 669 fits were executed, requiring 8 minutes and 52 seconds of user CPU time on a single core. This corresponds to an average of approximately 0.134 seconds per fit. All computations were carried out on a machine equipped with an Intel(R) Xeon(R) E5-2620 CPU running at 2.00 GHz.

\subsection{Extended tracks}\label{sec:extended-track}
We emphasize that even in the case of extended tracks, the reconstruction is carried out under the assumption of local straightness, in line with the current scope of the work. This choice ensures consistency with the nuclear-recoil topology of interest, while laying the groundwork for future extensions to more general track shapes.

The BAT fit infers the $(X, Y, L)$ parameters of the ionization tracks. This information is combined with the $\Delta Z$ component extracted from the analysis of PMT signals which, under the assumption of strictly straight tracks -- such as MIP-like particles, alpha particles, and nuclear recoils -- enables a 3D reconstruction. The $(X,Y)$ coordinates inferred from the BAT fit are then used to associate each PMT signal with a corresponding track in the camera image, where the transverse spatial resolution is significantly higher. 

Events producing extended ionization trails are first reduced to point-like tracklets through a slicing procedure. Each individual tracklet is then reconstructed using the BAT fit, as is done for localized interactions (see Section~\ref{sec:loc_track_rec}). Consequently, the computational cost for extended tracks is the same as that for localized ones, scaled by the number of slices.

When the PMT signal exhibits multiple peaks -- typical of MIP-like particles and extended ER -- a peak-finding algorithm is applied to identify the dominant features. Only peaks that are observed within the same time window by at least two PMTs, exceeding a defined threshold and  separated by a minimum of 60\,ns, are retained. Each selected peak is then treated as an independent localized event and reconstructed using the method described in Section~\ref{sec:loc_track_rec}. The $\Delta Z$ component of each reconstructed segment is obtained by measuring the time difference between consecutive peaks and converting it into a spatial distance using the electron drift velocity.

\begin{figure*}[th]
    \centering
    \begin{subfigure}{0.33\textwidth}
        \centering
        \raisebox{1.5mm}{\includegraphics[width=\textwidth]{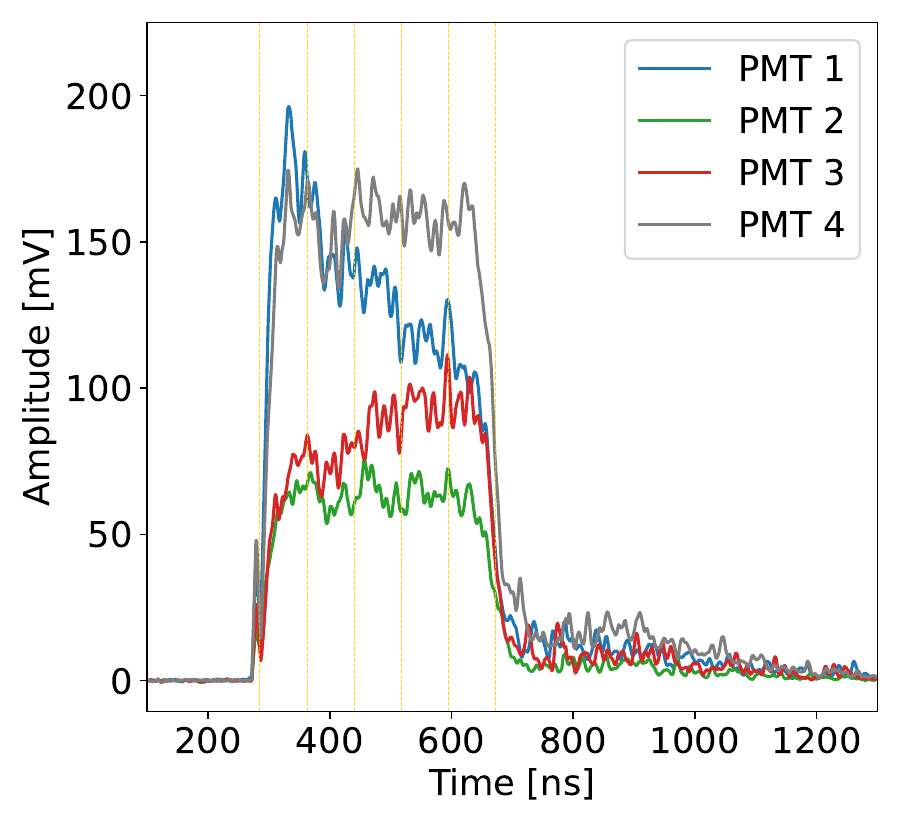}}
        \caption{}
        \label{fig:LIME:alpha_wf_sliced}
    \end{subfigure}\hfill
    \begin{subfigure}{0.33\textwidth}
        \centering
        \includegraphics[width=\textwidth]{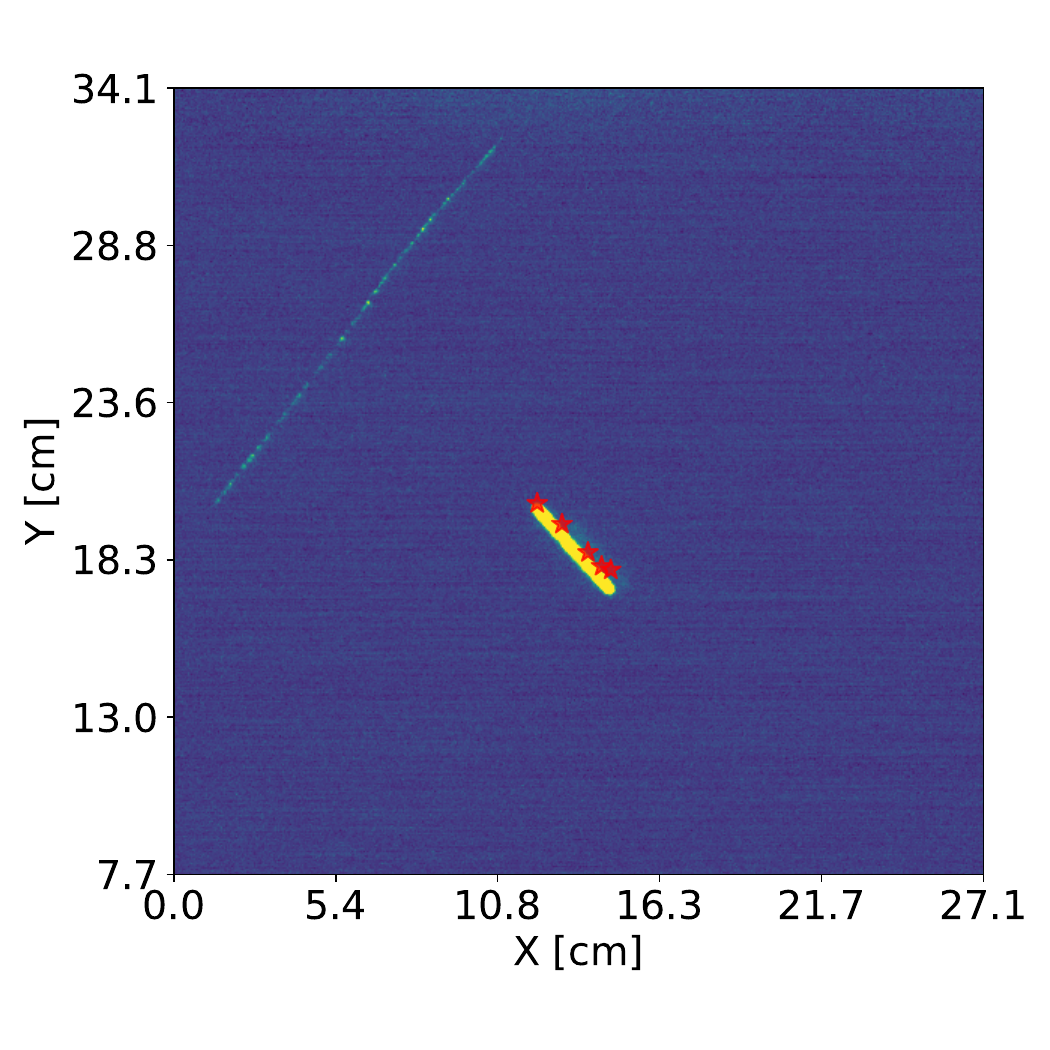}
        \caption{}
        \label{fig:LIME:batted_result}
    \end{subfigure}\hfill
    \begin{subfigure}{0.33\textwidth}
        \centering
        \includegraphics[width=\textwidth]{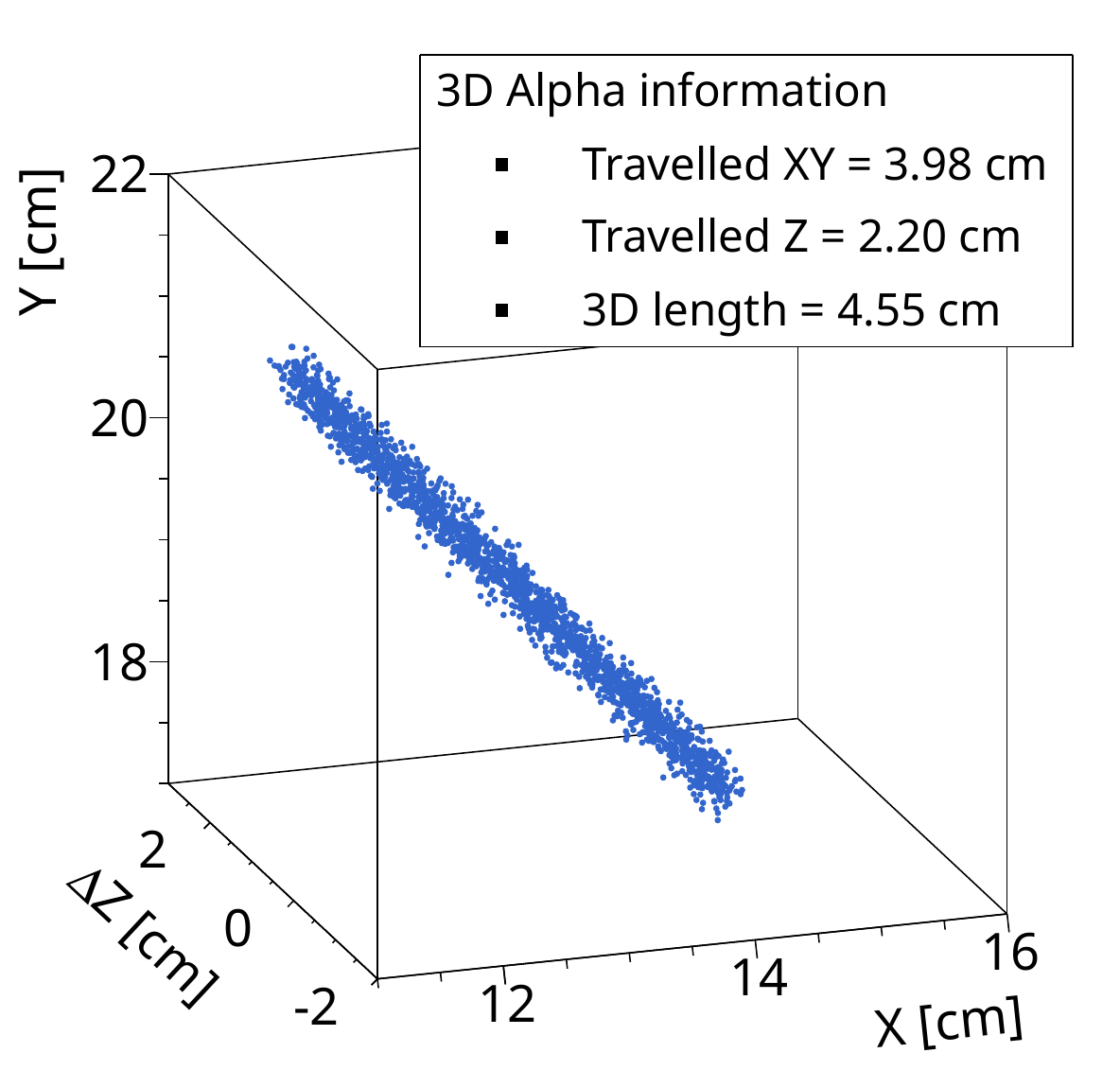}
        \caption{}
        \label{fig:alpha_3d}
    \end{subfigure}
   \caption{3D reconstruction of an extended alpha particle track. (\subref{fig:LIME:alpha_wf_sliced}) PMT signals with highlighted time windows used for the BAT fit; (\subref{fig:LIME:batted_result}) overlay of the BAT-reconstructed positions (red stars) on the corresponding camera image; (\subref{fig:alpha_3d}) final 3D representation of the alpha track combining PMT and camera information.}
    \label{fig:3D_on_alphas}
\end{figure*}

In the case of alpha particles, where the PMT signal appears as a continuous step-like signal without prominent peaks, a different strategy is adopted. The signal is divided into short time windows, each 60 ns wide, corresponding to the typical duration of a localized interaction. The integrated charge in each time slice is computed and fitted using the BAT algorithm, resulting in a set of $(X, Y, L)$ points. This information can be used either to match the PMT signals with the corresponding pixel cluster in the camera image, retrieving the $(XY)$ projection, or to directly reconstruct the 3D shape of the alpha track. Since individual peaks are absent in this case, the $\Delta Z$ coordinate is extracted by measuring the time difference between the last and the first slice.
An example of this procedure is shown in Fig.~\ref{fig:3D_on_alphas}, illustrating the key steps. In panel~\ref{fig:LIME:alpha_wf_sliced}, the alpha PMT signals are displayed, showing a high-amplitude, sustained signal with no prominent peaks. 

The multiple time windows used to segment the signal into short, localized interactions (60 ns each) are also indicated. This segmentation assumes straight tracks, under which condition a finer subdivision in $(x,y)$ would ultimately need to be recombined before matching with the PMT slices, yielding the same effective result. The resulting posterior $(X, Y)$ positions inferred with the BAT fit are shown as red stars in panel~\ref{fig:LIME:batted_result}, overlaid on the corresponding camera image. A clear match between the PMT-based and camera-based reconstructions is observed. The time-length for this track is measured to be 400\,ns, which corresponds to $\Delta Z$ = 2.2\,cm, assuming the nominal drift velocity. By combining the information from the PMT signals and the camera image, a 3D reconstruction of the alpha track is obtained, as shown in panel~\ref{fig:alpha_3d}, illustrating how the information from the two sensors can be merged into a single event display, which is ultimately also one of the objectives of CYGNO. For improved visualization, an ionization cloud is rendered by sampling random points from the transverse light distribution.

\begin{figure}[ht]
    \centering
    \includegraphics[width = 1\linewidth]{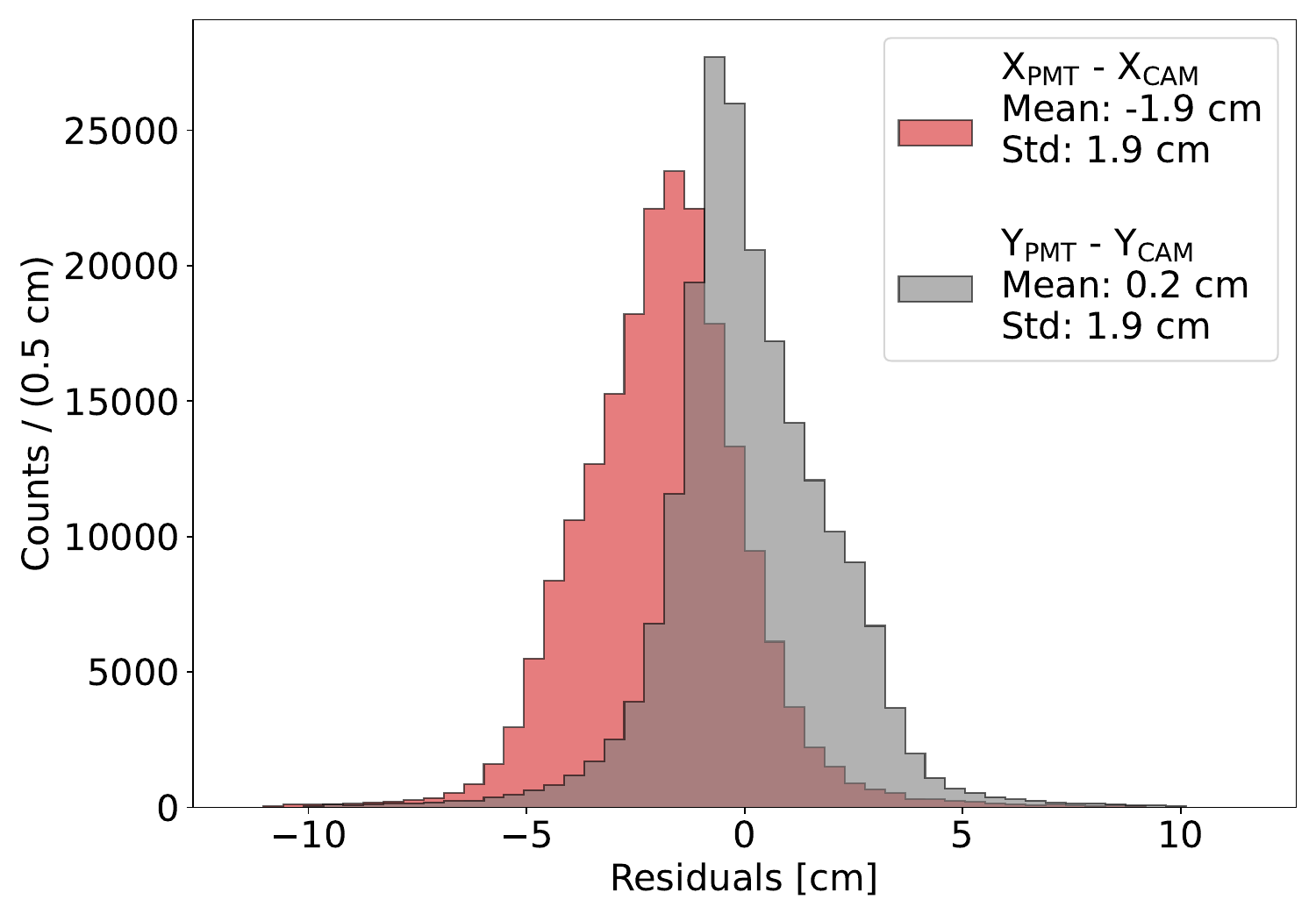}
    \caption{Distribution of the residuals $\Delta X$ and $\Delta Y$ between the PMT-based and camera-based track reconstructions for extended events. The tracks reconstructed from the camera images are resampled to match the number of points in the corresponding PMT waveform, and a point-by-point distance is computed.}
    \label{fig:BAT_distances}
\end{figure}

To quantify the matching between the PMT-based reconstruction and the track observed in the camera, the image track is resampled to the same number of points as the PMT waveform. Edge pixels are identified as start and end points, and intermediate points are uniformly distributed along the track in the $(XY)$ plane. A point-by-point distance is then computed between the two reconstructions. The residuals in $\Delta X$ and $\Delta Y$ are reported in Fig.~\ref{fig:BAT_distances}.  A systematic offset in the $X$ direction is observed, likely due to the distance-metric definition; further investigation is ongoing. 
Compared to the localized interactions of Fig.~\ref{fig:golden_association}, the residuals show a non Gaussian shape and a slightly larger spread of 1.9\,cm.
This is justified as the fit performance depends on the $(X,Y)$ position of the track and multiple correlated points per track contribute to the histograms. Event-by-event systematics, combined with the non-isotropic alpha emission in the detector, naturally lead to the observed non-Gaussian features and asymmetries. Overall, the reconstructed PMT tracks match the image projection in direction and topology. Since the camera provides no $Z$ information, the angular accuracy in three dimensions cannot be directly validated, but the consistency in $(XY)$ suggests comparable resolution along $Z$. This supports the robustness of the 3D reconstruction from PMT data alone.

The current model does not resolve the track sense (head–tail), as it assumes symmetric light emission. This limitation could be overcome by incorporating asymmetric light-yield models in future work. In a forthcoming paper we will extend the method to retrieve not only the 3D shape, but also the direction and sense of tracks -- crucial information for directional dark matter searches. This will also be applied to the study of alpha backgrounds in LIME, in particular radon progeny recoils.

\section{Conclusions} \label{sec:conclusions}
To our knowledge, this work provides the first demonstration that PMT signals alone can be used to reconstruct both the three-dimensional topology and the energy of ionization tracks in a gaseous TPC with optical readout. By modeling light collection probabilistically within the BAT framework and using MCMC sampling, we achieve robust in-situ calibration and event-by-event reconstruction. Validation with data from the CYGNO LIME prototype shows that both localized and extended tracks can be reconstructed with sub-centimeter precision. 

This represents a significant step forward for optical TPCs in rare-event searches. The ability to perform 3D reconstruction without a pixelated readout paves the way for scalable, cost-effective designs for directional dark matter detection. While the present implementation assumes straight tracks and neglects optical effects such as refraction at material boundaries, it already proves effective across diverse topologies. Future improvements will target low-light and overlapping signals, as well as asymmetric light-yield models to recover track direction and sense, crucial for identifying WIMP-induced nuclear recoils.

These techniques are currently being applied to the study of alpha-induced backgrounds in LIME, including radon progeny recoils, and will be detailed in forthcoming work. In parallel, the results reported here have informed the optimization of PMT layout for {\tt CYGNO-04}, a $\mathcal{O}$(1\,m$^3$) detector under design.

In conclusion, the PMT-only reconstruction with BAT achieves a sub-centimeter spatial resolutions and an energy resolutions comparable to that of the sCMOS, enables robust event matching even under pile-up, and provides a solid basis for 3D directional sensitivity, an essential capability for future dark matter searches.

\section*{Acknowledgements}

This project has received fundings under the European Union’s Horizon 2020 research and innovation program from the European Research Council (ERC) grant agreement No. 818744 and is supported by the Italian Ministry of Education, University and Research through the project PRIN: Progetti di Ricerca di Rilevante Interesse Nazionale “Zero Radioactivity in Future experiment” (Prot. 2017T54J9J). We want to thank General Services and Mechanical Workshops of Laboratori Nazionali di Frascati (LNF). We want to thank the INFN Laboratori Nazionali del Gran Sasso for hosting and supporting the Cygno project.


\bibliographystyle{spphys}
\bibliography{Reference.bib}
\end{document}

%% file: author_list_EPJC-new.tex
\author{
Fernando Domingues Amaro\inst{1} %
           \and
Rita Antonietti\inst{2,3} %
           \and
Elisabetta Baracchini \inst{4,5} %
           \and
Luigi  Benussi \inst{6} %
           \and
Stefano Bianco \inst{6} %
            \and    
Francesco Borra \inst{2,3}\thanks{\textit{corresponding author:} francesco.borra@uniroma3.it}
          \and
Cesidio Capoccia \inst{6} %
          \and
Michele Caponero \inst{6,9} %
           \and
Gianluca Cavoto \inst{7,8} %
           \and
Igor Abritta Costa \inst{6} %
           \and
Antonio Croce \inst{6} %
           \and           
Emiliano Dan\'e \inst{6} %
           \and
Melba D'Astolfo \inst{4,5} %
           \and
Giorgio Dho \inst{6} %
           \and
Flaminia Di Giambattista \inst{4,5} %
           \and
Emanuele Di Marco \inst{7} %
          \and
Giulia D'Imperio \inst{7} %
          \and
Matteo Folcarelli \inst{7,8}\thanks{\textit{corresponding author:} matteo.folcarelli@uniroma1.it}
           \and
Joaquim Marques Ferreira dos Santos \inst{1} %
           \and
Davide Fiorina \inst{4,5} %
           \and
Francesco Iacoangeli \inst{7} %
           \and
Zahoor Ul Islam \inst{4,5} %
           \and
Herman Pessoa Lima J\'unior \inst{4,5} %
           \and
Ernesto Kemp\inst{10}
           \and
Giovanni Maccarrone \inst{6} %
           \and
Rui Daniel Passos Mano \inst{1} %
           \and
David Jos\'e Gaspar  Marques \inst{4,5}\thanks{\textit{corresponding author:} david.marques@gssi.it} %
           \and
Luan Gomes Mattosinhos de Carvalho \inst{12} %
           \and
Giovanni Mazzitelli \inst{6} %
           \and
Alasdair Gregor McLean \inst{11} %
          \and
Pietro Meloni\inst{2,3} %
           \and
Andrea Messina \inst{7,8} %
           \and
Cristina Maria Bernardes Monteiro \inst{1} %
           \and
Rafael Antunes Nobrega \inst{12} %
           \and
Igor Fonseca Pains \inst{12} %
          \and
Emiliano Paoletti \inst{6} %
           \and
Luciano Passamonti \inst{6} %
           \and
Fabrizio Petrucci \inst{2,3} %
           \and
Stefano Piacentini \inst{4,5} %
           \and
Davide Piccolo \inst{6} %
           \and
Daniele Pierluigi \inst{6} %
           \and
Davide Pinci \inst{7}
           \and
Atul Prajapati \inst{4,5}\thanks{\textit{Currently at:} University of L'Aquila, Edificio Renato Ricamo, via Vetoio, Coppito - 67100 L'Aquila, Italy} %
           \and
Francesco Renga \inst{7} %
           \and
Rita Joana Cruz Roque \inst{1} %
           \and
Filippo Rosatelli \inst{6} %
           \and
Alessandro Russo \inst{6} %
          \and
Giovanna Saviano \inst{6,13} %
          \and
Pedro Alberto Oliveira Costa Silva \inst{1}
           \and
Neil John Curwen Spooner \inst{11}
           \and
Roberto Tesauro \inst{6}
           \and
Sandro Tomassini \inst{6}
           \and
Samuele Torelli \inst{4,5}\thanks{\textit{Currently at:} Donostia International Physics Center, BERC Basque Excellence Research Centre, Manuel Lardizabal 4, San Sebastián / Donostia, E-20018, Spain}
           \and
Donatella Tozzi \inst{7,8}
}


\institute{LIBPhys, Department of Physics, University of Coimbra, 3004-516 Coimbra, Portugal; \label{1} 
           \and
Istituto Nazionale di Fisica Nucleare, Sezione di Roma TRE, 00146, Roma, Italy; \label{2}
            \and
 Dipartimento di Matematica e Fisica, Universit\`a Roma TRE, 00146, Roma, Italy; \label{3}
            \and
Gran Sasso Science Institute, 67100, L'Aquila, Italy; \label{4}
           \and
 Istituto Nazionale di Fisica Nucleare, Laboratori Nazionali del Gran Sasso, 67100, Assergi, Italy; \label{5}
            \and
 Istituto Nazionale di Fisica Nucleare, Laboratori Nazionali  di Frascati,  00044, Frascati, Italy; \label{6}
           \and
Istituto Nazionale di Fisica Nucleare, Sezione di Roma, 00185, Rome, Italy; \label{7}
           \and
  Dipartimento di Fisica, Sapienza Universit\`a di Roma, 00185, Roma, Italy; \label{8}
           \and
 ENEA Centro Ricerche Frascati, 00044, Frascati, Italy; \label{9}
           \and
Universidade Estadual de Campinas  - UNICAMP,  Campinas 13083-859, SP, Brazil; \label{10}
           \and
 Department of Physics and Astronomy, University of Sheffield, Sheffield, S3 7RH, UK; \label{11}
 \and
  Universidade Federal de Juiz de Fora, Faculdade de Engenharia, 36036-900, Juiz de Fora, MG, Brasil; \label{12}
           \and
 Dipartimento di Ingegneria Chimica, Materiali e Ambiente, Sapienza Universit\`a di Roma, 00185, Roma, Italy; \label{13}
}

%% file: Reference.bib
@article{Bertone:2004pz,
title = {Particle dark matter: evidence, candidates and constraints},
journal = {Physics Reports},
volume = {405},
number = {5},
pages = {279-390},
year = {2005},
issn = {0370-1573},
doi = {https://doi.org/10.1016/j.physrep.2004.08.031},
url = {https://www.sciencedirect.com/science/article/pii/S0370157304003515},
author = {Gianfranco Bertone and et al.}
}

@article{Green:2006cb,
    author = "Green, Anne M. and Morgan, Ben",
    title = "{Optimizing WIMP directional detectors}",
    eprint = "astro-ph/0609115",
    archivePrefix = "arXiv",
    doi = "10.1016/j.astropartphys.2006.10.006",
    journal = "Astropart. Phys.",
    volume = "27",
    pages = "142--149",
    year = "2007"
}

@article{Mayet_2016zxu,
	doi = {10.1016/j.physrep.2016.02.007},
	
	url = {https://doi.org/10.1016\%2Fj.physrep.2016.02.007},
	
	year = 2016,
	month = {apr},
	
	publisher = {Elsevier {BV}
	},
	
	volume = {627},
	
	pages = {1--49},
	author = {F. Mayet and et al},	
	title = {A review of the discovery reach of directional Dark Matter detection},	
	journal = {Physics Reports}
}

@article{Amaro:2022gub,
	doi = {10.3390/instruments6010006},
	
	url = {https://doi.org/10.3390\%2Finstruments6010006},
	
	year = 2022,
	month = {jan},
	
	publisher = {{MDPI} {AG}
	},
	
	volume = {6},
	
	number = {1},
	
	pages = {6},
	
	author = {Fernando Domingues Amaro and Elisabetta Baracchini and et al.},
	
	title = {The {CYGNO} Experiment},
	
	journal = {Instruments}
}

@article{bib:Fraga:2003uu,
	title = {The GEM scintillation in He–CF4, Ar–CF4, Ar–TEA and Xe–TEA mixtures},
	journal = {Nuclear Instruments and Methods in Physics Research Section A: Accelerators, Spectrometers, Detectors and Associated Equipment},
	volume = {504},
	number = {1},
	pages = {88-92},
	year = {2003},
	note = {Proceedings of the 3rd International Conference on New Developments in Photodetection},
	issn = {0168-9002},
	doi = {https://doi.org/10.1016/S0168-9002(03)00758-7},
	url = {https://www.sciencedirect.com/science/article/pii/S0168900203007587},
	author = {M.M.F.R. Fraga and F.A.F. Fraga and S.T.G. Fetal and L.M.S. Margato and R.Ferreira Marques and A.J.P.L. Policarpo}
}

@article{Miernik:2007cwz,
    author = "Miernik, K. and others",
    editor = "Ferreira, Lidia S. and Arumugam, Paramasivan",
    title = "{Imaging nuclear decays with Optical Time Projection Chamber}",
    doi = "10.1063/1.2827275",
    journal = "AIP Conf. Proc.",
    volume = "961",
    number = "1",
    pages = "307",
    year = "2007"
}

@article{Antochi:2018otx,
    author = "Antochi, V. C. and Baracchini, E. and Cavoto, G. and Marco, E. Di and Marafini, M. and Mazzitelli, G. and Pinci, D. and Renga, F. and Tomassini, S. and Voena, C.",
    title = "{Combined readout of a triple-GEM detector}",
    eprint = "1803.06860",
    archivePrefix = "arXiv",
    primaryClass = "physics.ins-det",
    doi = "10.1088/1748-0221/13/05/P05001",
    journal = "JINST",
    volume = "13",
    number = "05",
    pages = "P05001",
    year = "2018"
}

@article{bib:fe55,
	doi = {10.1088/1748-0221/14/07/p07011},
	
	url = {https://doi.org/10.1088\%2F1748-0221%2F14%2F07%2Fp07011},
	
	year = 2019,
	month = {jul},
	
	publisher = {{IOP} Publishing},
	
	volume = {14},
	
	number = {07},
	
	pages = {P07011--P07011},
	
	author = {I. Abritta Costa and et al},
	
	title = {Performance of optically readout {GEM}-based {TPC} with a $^{55}$Fe source},
	
	journal = {Journal of Instrumentation}
}

@article{LIME,
  author    = {F. D. Amaro and R. Antonietti and E. Baracchini and others},
  title     = {A 50 l Cygno prototype overground characterization},
  journal   = {The European Physical Journal C},
  volume    = {83},
  pages     = {946},
  year      = {2023},
  doi       = {10.1140/epjc/s10052-023-11988-9}
}

@book{optics,
  title={Introduction to Optics},
  isbn={9788131720240},
  url={https://books.google.it/books?id=-XjoZQVxoNIC},
  year={2008},
  publisher={Pearson Education}
}

@article{Baracchini_2021,
doi = {10.1088/1361-6501/abbd12},
url = {https://dx.doi.org/10.1088/1361-6501/abbd12},
year = {2020},
month = {dec},
publisher = {IOP Publishing},
volume = {32},
number = {2},
pages = {025902},
author = {Baracchini, E and others},
title = {Identification of low energy nuclear recoils in a gas time projection chamber with optical readout},
journal = {Measurement Science and Technology}
}

@book{pearlbook,
author = "Pearl, Judea",
title = "{Probabilistic Reasoning in Intelligent Systems: Networks of Plausible Inference}",
publisher = "Morgan-Kaufmann",
year = "1988",
}

@book{jensenbook,
author = "Jensen, F. V.",
title = "{An Introduction to Bayesian Networks}",
publisher = "UCL Press",
year = "1996",
}

@book{kollerbook,
author = "Koller, D. and Friedman, N.",
title = "{Probabilistic Graphical Models: Principles and Techniques}",
publisher = "MIT Press",
year  = "2009",
}

@book{DAgostini:2003syq,
    author = "D'Agostini, G.",
    title = "{Bayesian reasoning in data analysis: A critical introduction}",
    publisher = "{World Industries Scientific Publishing}",
    ISBN = "{9789814447959}",
    year = "{2003}"
}

@misc{ref:BAT,
    author = "F. Beaujean and A. Caldwell and D. Greenwald and D. Kollar and K. Kröninger and O. Schulz",
    title = "{Bayesian Analysis Toolkit: BAT}",
	Howpublished = {\url{https://github.com/bat/bat}},
    url = {https://github.com/bat/bat.}
}

@ARTICLE{2009CoPhC.180.2197C,
       author = {{Caldwell}, Allen and {Koll{\'a}r}, Daniel and {Kr{\"o}ninger}, Kevin},
        title = "{BAT - The Bayesian analysis toolkit}",
      journal = {Computer Physics Communications},
         year = 2009,
        month = nov,
       volume = {180},
       number = {11},
        pages = {2197-2209},
          doi = {10.1016/j.cpc.2009.06.026},
archivePrefix = {arXiv},
       eprint = {0808.2552},
 primaryClass = {physics.data-an},
       adsurl = {https://ui.adsabs.harvard.edu/abs/2009CoPhC.180.2197C}
}

@article{Ciuchini:2000de,
    author = "Ciuchini, Marco and D'Agostini, G. and Franco, E. and Lubicz, V. and Martinelli, G. and Parodi, F. and Roudeau, P. and Stocchi, A.",
    title = "{2000 CKM triangle analysis: A Critical review with updated experimental inputs and theoretical parameters}",
    eprint = "hep-ph/0012308",
    archivePrefix = "arXiv",
    reportNumber = "LAL-00-77, ROME1-1307-00, RM3-TH-00-16",
    doi = "10.1088/1126-6708/2001/07/013",
    journal = "JHEP",
    volume = "07",
    pages = "013",
    year = "2001"
}

@article{deBlas:2021wap,
    author = "de Blas, J. and Ciuchini, M. and Franco, E. and Goncalves, A. and Mishima, S. and Pierini, M. and Reina, L. and Silvestrini, L.",
    title = "{Global analysis of electroweak data in the Standard Model}",
    eprint = "2112.07274",
    archivePrefix = "arXiv",
    primaryClass = "hep-ph",
    reportNumber = "KEK-TH-2378",
    doi = "10.1103/PhysRevD.106.033003",
    journal = "Phys. Rev. D",
    volume = "106",
    number = "3",
    pages = "033003",
    year = "2022"
}

@article{Sauli1997,
  author    = {F. Sauli},
  title     = {GEM: A new concept for electron amplification in gas detectors},
  journal   = {Nucl. Instrum. Meth. A},
  volume    = {386},
  year      = {1997},
  pages     = {531--534},
  doi       = {10.1016/S0168-9002(96)01172-2}
}

@misc{Vahsen:2020pzb,
  author = {S.E. Vahsen and others},
  title  = {CYGNUS: Feasibility of a nuclear recoil observatory with directional sensitivity %to dark matter and neutrinos},
  year   = {2020},
  note   = {arXiv:\,2008.12587\, [physics.ins-det]}
}

@article{DarkSide-50:2023fcw,
    author = "Agnes, P. and others",
    collaboration = "DarkSide-50",
    title = "{Search for low mass dark matter in DarkSide-50: the bayesian network approach}",
    eprint = "2302.01830",
    archivePrefix = "arXiv",
    primaryClass = "hep-ex",
    reportNumber = "FERMILAB-PUB-23-043-AD-CSAID-ND",
    doi = "10.1140/epjc/s10052-023-11410-4",
    journal = "Eur. Phys. J. C",
    volume = "83",
    pages = "322",
    year = "2023"
}

@article{XENONCollaboration:2023dar,
    author = "Aprile, E. and others",
    collaboration = "(XENON Collaboration)\textdagger{}\textdagger{}, XENON",
    title = "{Detector signal characterization with a Bayesian network in XENONnT}",
    eprint = "2304.05428",
    archivePrefix = "arXiv",
    primaryClass = "hep-ex",
    doi = "10.1103/PhysRevD.108.012016",
    journal = "Phys. Rev. D",
    volume = "108",
    number = "1",
    pages = "012016",
    year = "2023"
}

@article{Deaconu:2017vam,
    author = "Deaconu, Cosmin and Leyton, Michael and Corliss, Ross and Druitt, Gabriela and Eggleston, Richard and Guerrero, Natalia and Henderson, Shawn and Lopez, Jeremy and Monroe, Jocelyn and Fisher, Peter",
    title = "{Measurement of the directional sensitivity of Dark Matter Time Projection Chamber detectors}",
    eprint = "1705.05965",
    archivePrefix = "arXiv",
    primaryClass = "astro-ph.IM",
    doi = "10.1103/PhysRevD.95.122002",
    journal = "Phys. Rev. D",
    volume = "95",
    number = "12",
    pages = "122002",
    year = "2017"
}

@article{XENON:2019zpr,
    author = "Aprile, E. and others",
    collaboration = "XENON",
    title = "{Search for Light Dark Matter Interactions Enhanced by the Migdal Effect or Bremsstrahlung in XENON1T}",
    eprint = "1907.12771",
    archivePrefix = "arXiv",
    primaryClass = "hep-ex",
    doi = "10.1103/PhysRevLett.123.241803",
    journal = "Phys. Rev. Lett.",
    volume = "123",
    number = "24",
    pages = "241803",
    year = "2019"
}

@article{DarkSide:2018bpj,
    author = "Agnes, P. and others",
    collaboration = "DarkSide",
    title = "{Low-Mass Dark Matter Search with the DarkSide-50 Experiment}",
    eprint = "1802.06994",
    archivePrefix = "arXiv",
    primaryClass = "astro-ph.HE",
    reportNumber = "FERMILAB-PUB-18-048-AE-E-PPD",
    doi = "10.1103/PhysRevLett.121.081307",
    journal = "Phys. Rev. Lett.",
    volume = "121",
    number = "8",
    pages = "081307",
    year = "2018"
}

@article{DarkSide:2018ppu,
    author = "Agnes, P. and others",
    collaboration = "DarkSide",
    title = "{Constraints on Sub-GeV Dark-Matter\textendash{}Electron Scattering from the DarkSide-50 Experiment}",
    eprint = "1802.06998",
    archivePrefix = "arXiv",
    primaryClass = "astro-ph.CO",
    reportNumber = "FERMILAB-PUB-18-052-AD-AE-CD-E",
    doi = "10.1103/PhysRevLett.121.111303",
    journal = "Phys. Rev. Lett.",
    volume = "121",
    number = "11",
    pages = "111303",
    year = "2018"
}

@article{XENON:2019gfn,
    author = "Aprile, E. and others",
    collaboration = "XENON",
    title = "{Light Dark Matter Search with Ionization Signals in XENON1T}",
    eprint = "1907.11485",
    archivePrefix = "arXiv",
    primaryClass = "hep-ex",
    doi = "10.1103/PhysRevLett.123.251801",
    journal = "Phys. Rev. Lett.",
    volume = "123",
    number = "25",
    pages = "251801",
    year = "2019"
}

@article{DarkSide:2018kuk,
    author = "Agnes, P. and others",
    collaboration = "DarkSide",
    title = "{DarkSide-50 532-day Dark Matter Search with Low-Radioactivity Argon}",
    eprint = "1802.07198",
    archivePrefix = "arXiv",
    primaryClass = "astro-ph.CO",
    reportNumber = "FERMILAB-PUB-18-053-AD-AE-CD-E",
    doi = "10.1103/PhysRevD.98.102006",
    journal = "Phys. Rev. D",
    volume = "98",
    number = "10",
    pages = "102006",
    year = "2018"
}

@article{XENON:2016jmt,
    author = "Aprile, E. and others",
    collaboration = "XENON",
    title = "{Low-mass dark matter search using ionization signals in XENON100}",
    eprint = "1605.06262",
    archivePrefix = "arXiv",
    primaryClass = "astro-ph.CO",
    doi = "10.1103/PhysRevD.94.092001",
    journal = "Phys. Rev. D",
    volume = "94",
    number = "9",
    pages = "092001",
    year = "2016",
    note = "[Erratum: Phys.Rev.D 95, 059901 (2017)]"
}

@article{SuperCDMS:2024yiv,
    author = "Albakry, M. F. and others",
    collaboration = "SuperCDMS",
    title = "{Light dark matter constraints from SuperCDMS HVeV detectors operated underground with an anticoincidence event selection}",
    eprint = "2407.08085",
    archivePrefix = "arXiv",
    primaryClass = "hep-ex",
    reportNumber = "FERMILAB-PUB-24-0376-PPD",
    doi = "10.1103/PhysRevD.111.012006",
    journal = "Phys. Rev. D",
    volume = "111",
    number = "1",
    pages = "012006",
    year = "2025"
}

@article{DEAPCollaboration:2021raj,
    author = "Adhikari, P. and others",
    collaboration = "(DEAP Collaboration)\textdaggerdbl{}, DEAP",
    title = "{First Direct Detection Constraints on Planck-Scale Mass Dark Matter with Multiple-Scatter Signatures Using the DEAP-3600 Detector}",
    eprint = "2108.09405",
    archivePrefix = "arXiv",
    primaryClass = "astro-ph.CO",
    doi = "10.1103/PhysRevLett.128.011801",
    journal = "Phys. Rev. Lett.",
    volume = "128",
    number = "1",
    pages = "011801",
    year = "2022"
}

@article{ParticleDataGroup:2024cfk,
    author = "Navas, S. and others",
    collaboration = "Particle Data Group",
    title = "{Review of particle physics}",
    doi = "10.1103/PhysRevD.110.030001",
    journal = "Phys. Rev. D",
    volume = "110",
    number = "3",
    pages = "030001",
    year = "2024"
}

@article{FRAGA200388_rydberg_state,
title = {The GEM scintillation in He–CF4, Ar–CF4, Ar–TEA and Xe–TEA mixtures},
journal = {Nuclear Instruments and Methods in Physics Research Section A: Accelerators, Spectrometers, Detectors and Associated Equipment},
volume = {504},
number = {1},
pages = {88-92},
year = {2003},
note = {Proceedings of the 3rd International Conference on New Developments in Photodetection},
issn = {0168-9002},
doi = {https://doi.org/10.1016/S0168-9002(03)00758-7},
url = {https://www.sciencedirect.com/science/article/pii/S0168900203007587},
author = {M.M.F.R. Fraga and F.A.F. Fraga and S.T.G. Fetal and L.M.S. Margato and R.Ferreira Marques and A.J.P.L. Policarpo}
}

@article{ELY,
    author = "Amaro, Fernando Domingues and others",
    title = "{Enhancing the light yield of He:CF$_4$ based gaseous detector}",
    eprint = "2406.05713",
    archivePrefix = "arXiv",
    primaryClass = "physics.ins-det",
    doi = "10.1140/epjc/s10052-024-13471-5",
    journal = "Eur. Phys. J. C",
    volume = "84",
    number = "10",
    pages = "1122",
    year = "2024"
}

@misc{pmt_handbook,
  title = {Photomultiplier Tubes: Basics and Applications, 3rd Edition},
  howpublished = {\url{https://www.hamamatsu.com/content/dam/hamamatsu-photonics/sites/documents/99_SALES_LIBRARY/etd/PMT_handbook_v4E.pdf}},
  note = {Accessed: 2025-02-04}
}

@article{NEWAGE2015,
  author  = {Nakamura, Kiseki and Miuchi, Kentaro and Tanimori, Toru and Kubo, Hidetoshi and Takada, Atsushi and Parker, Joseph D. and Mizumoto, Tetsuya and Mizumura, Yoshitaka and Nishimura, Hironobu and Sekiya, Hiroyuki and Takeda, Atsushi},
  title   = {Direction-sensitive dark matter search with gaseous tracking detector NEWAGE-0.3b'},
  journal = {Progress of Theoretical and Experimental Physics},
  year    = {2015},
  number  = {4},
  pages   = {043F01},
  doi     = {10.1093/ptep/ptv041}
}

@article{MIMAC2011,
  author  = {Santos, D. and et al.},
  title   = {MIMAC: A micro-TPC matrix for directional detection of dark matter},
  journal = {Journal of Physics: Conference Series},
  volume  = {309},
  pages   = {012014},
  year    = {2011},
  doi     = {10.1088/1742-6596/309/1/012014}
}

@article{DMTPC2011,
  author  = {Ahlen, S. and et al.},
  title   = {First Dark Matter Search Results from a Surface Run of the 10-L DMTPC Directional Dark Matter Detector},
  journal = {Physics Letters B},
  volume  = {695},
  pages   = {124--129},
  year    = {2011},
  doi     = {10.1016/j.physletb.2010.11.041}
}

@article{DRIFT2012,
  author  = {Daw, E. and et al.},
  title   = {Spin-dependent limits from the DRIFT-IId directional dark matter detector},
  journal = {Astroparticle Physics},
  volume  = {35},
  number  = {7},
  pages   = {397--401},
  year    = {2012},
  doi     = {10.1016/j.astropartphys.2011.11.003}
}

@article{DarkSide20k:2025tes,
    author = "Ave, M. and others",
    collaboration = "DarkSide 20k",
    title = "{Characterization of low energy argon recoils with ReD and ReD+}",
    doi = "10.1088/1748-0221/20/06/C06062",
    journal = "JINST",
    volume = "20",
    number = "06",
    pages = "C06062",
    year = "2025"
}

@article{Bondar2017,
  author    = {Bondar, A. and Buzulutskov, A. and Cortesi, M. and Dolgoshein, B. and Kimura, M. and et al.},
  title     = {Study of columnar recombination in liquid argon},
  journal   = {Journal of Instrumentation},
  volume    = {12},
  number    = {06},
  pages     = {C06018},
  year      = {2017},
  doi       = {10.1088/1748-0221/12/06/C06018}
}

@article{Lisotti:2024fco,
    author = "Lisotti, Chiara and others",
    title = "{CYG$\nu $S: detecting solar neutrinos with directional gas time projection chambers}",
    eprint = "2404.03690",
    archivePrefix = "arXiv",
    primaryClass = "hep-ph",
    doi = "10.1140/epjc/s10052-024-13392-3",
    journal = "Eur. Phys. J. C",
    volume = "84",
    number = "10",
    pages = "1021",
    year = "2024"
}

@article{NEXT:2012rto,
title = {Initial results of NEXT-DEMO, a large-scale prototype of the NEXT-100 experiment},
author = {V. Alvarez and others},
doi = {10.1088/1748-0221/8/04/P04002},
year = {2013},
date = {2013-01-01},
journal = {JINST},
volume = {8},
pages = {P04002},
pubstate = {published},
tppubtype = {article}
}

@article{Veenhof:1998tt,
    author = "Veenhof, R.",
    editor = "Krammer, M. and Neuhofer, G. and Regler, M. and Taurok, A.",
    title = "{GARFIELD, recent developments}",
    doi = "10.1016/S0168-9002(98)00851-1",
    journal = "Nucl. Instrum. Meth. A",
    volume = "419",
    pages = "726--730",
    year = "1998"
}

@article{Baracchini:2020phn,
    author = "Baracchini, E. and others",
    title = "{Stability and detection performance of a GEM-based Optical Readout TPC with He/CF$_4$ gas mixtures}",
    eprint = "2007.00608",
    archivePrefix = "arXiv",
    primaryClass = "physics.ins-det",
    doi = "10.1088/1748-0221/15/10/P10001",
    journal = "JINST",
    volume = "15",
    number = "10",
    pages = "P10001",
    year = "2020"
}
